\documentclass[npg]{copernicus}
\usepackage{amsmath}
\usepackage{graphicx}
\usepackage{ifthen}
\usepackage{amsfonts}
\usepackage{amssymb}
\usepackage{epsfig}
\usepackage{url}

\makeatletter
\def\cF{{\mathcal{F}}}

\def\cG{{\mathcal{G}}}

\def\PP{{\mathcal{P}}}

\def\be{\begin{equation}}
\def\ee{\end{equation}}
\newtheorem{theorem}{Theorem}

\newtheorem{proposition}[theorem]{Proposition}

\begin{document}
\title{A delay differential model of ENSO variability:\\
Parametric instability and the distribution of extremes}
\author[1]{Michael Ghil}
\author[2]{Ilya Zaliapin}
\author[3]{Sylvester Thompson}

\affil[1]{D\'{e}partement Terre-Atmosph\`{e}re-Oc\'ean and
Laboratoire de M\'{e}t\'{e}orologie Dynamique,
Ecole Normale Sup\'{e}rieure, Paris, FRANCE and
Department of Atmospheric and Oceanic Sciences and
Institute of Geophysics and Planetary Physics,
University of California Los Angeles, USA.
E-mail: ghil@atmos.ucla.edu;}
\affil[2]{Department of Mathematics and Statistics,
University of Nevada, Reno, USA.
E-mail: zal@unr.edu;}
\affil[3]{Department of Mathematics and Statistics,
University of Radford, Virginia, USA.
E-mail: thompson@radford.edu.}

\runningtitle{A delay differential model of ENSO variability}

\runningauthor{M. Ghil {\it et al.}}

\correspondence{Michael Ghil (ghil@atmos.ucla.edu)}

\received{}
\pubdiscuss{} 
\revised{}
\accepted{}
\published{}

\firstpage{1}

\maketitle

\begin{abstract}
We consider a delay differential equation (DDE) model for El-Niñ\~no Southern 
Oscillation (ENSO) variability. 
The model combines two key mechanisms that participate in ENSO dynamics: 
delayed negative feedback and seasonal forcing.
We perform stability analyses of the model in the three-dimensional space of its 
physically relevant parameters. 
Our results illustrate the role of these three parameters:
strength of seasonal forcing $b$, atmosphere-ocean coupling $\kappa$, 
and propagation period $\tau$ of oceanic waves across the Tropical Pacific.
Two regimes of variability, stable and unstable, are separated by a sharp neutral 
curve in the $(b,\tau)$ plane at constant $\kappa$.
The detailed structure of the neutral curve becomes very irregular and
possibly fractal, while individual trajectories within the unstable region
become highly complex and possibly chaotic, as the atmosphere-ocean coupling 
$\kappa$ increases.
In the unstable regime, spontaneous transitions occur in the mean ``temperature'' 
({\it i.e.}, thermocline depth), period, and extreme annual values, 
for purely periodic, seasonal forcing.
The model reproduces the Devil's bleachers characterizing other ENSO models, 
such as nonlinear, coupled systems of partial differential equations;
some of the features of this behavior have been documented in general circulation
models, as well as in observations.  
We expect, therefore, similar behavior in much more detailed and realistic models, 
where it is harder to describe its causes as completely.
\end{abstract}

\introduction[Introduction and motivation]

\subsection{Key ingredients of ENSO theory}
The El-Ni\~no/Southern-Oscillation (ENSO) phenomenon is the most 
prominent signal of seasonal-to-interannual climate variability. 
It was known for centuries to fishermen along the west coast of 
South America, who witnessed a seemingly sporadic and abrupt warming 
of the cold, nutrient-rich waters that support the food chain in those 
regions; these warmings caused havoc to their fish harvests \citep{Diaz92,Phil90}. 
The common occurrence of such warming shortly after Christmas inspired them to 
name it El Ni\~no, after the ``Christ child.'' 
Starting in the 1970s, El Ni\~no's climatic effects were found to 
be far broader than just its manifestations off the shores of Peru
\citep{Diaz92,Glantz+91}.
This realization led to a global awareness of ENSO's significance, 
and an impetus to attempt and improve predictions 
of exceptionally strong El Ni\~no events \citep{Latif94}.

The following conceptual elements have been shown to play a determining role
in the dynamics of the ENSO phenomenon.

{\bf \textit{(i) The Bjerknes hypothesis:}} 
\citet{Bjer69}, who laid the foundation of modern ENSO research, 
suggested a {\em {positive feedback}} as a mechanism for the growth of an 
internal instability that could produce large positive anomalies of sea surface
temperatures (SSTs) in the eastern Tropical Pacific.
We use here the climatological meaning of the term {\em anomaly},
{\em i.e.}, the difference between an instantaneous (or short-term average)
value and the {\em normal} (or long-term mean).
Using observations from the International Geophysical Year (1957-1958), Bjerknes 
realized that this mechanism must involve {\em{air-sea interaction}} in the tropics. 
The ``chain reaction'' starts with an initial warming of SSTs in the ``cold tongue'' 
that occupies the eastern part of the equatorial Pacific. 
This warming causes a weakening of the thermally direct Walker-cell 
circulation; this circulation involves air rising over the
warmer SSTs near Indonesia and sinking over the colder SSTs near Peru.
As the trade winds blowing from the east weaken and thus give way to 
westerly wind anomalies, the ensuing local changes in the ocean 
circulation encourage further SST increase. 
Thus the feedback loop is closed and further amplification of the 
instability is triggered.

{\bf \textit{(ii) Delayed oceanic wave adjustments:}} 
Compensating for Bjerknes's positive feedback is a {\em{negative feedback}} in the system 
that allows a return to colder conditions in the basin's eastern part \citep{SS88}. 
During the peak of the cold-tongue warming, called the 
{\em warm} or {\em El~Ni\~no} phase of ENSO, westerly wind anomalies 
prevail in the central part of the basin. 
As part of the ocean's adjustment to this atmospheric forcing, 
a Kelvin wave is set up in the tropical wave guide and carries 
a warming signal eastward; 
this signal deepens the eastern-basin thermocline,
which separates the warmer, well-mixed surface waters from the 
colder waters below, and thus contributes to the positive feedback 
described above. 
Concurrently, slower Rossby waves propagate westward, 
and are reflected at the basin's western boundary, giving rise 
therewith to an eastward-propagating Kelvin wave that has a cooling, 
thermocline-shoaling effect. 
Over time, the arrival of this signal erodes the warm event, 
ultimately causing a switch to a {\em cold}, {\em La~Ni\~na} phase.

{\bf \textit{(iii) Seasonal forcing:} } 
A growing body of work \citep{GR00,Cha94,Cha95,JNG94,JNG96,Tzi+94,Tzi+95} 
points to resonances between the Pacific basin's intrinsic air-sea 
oscillator and the annual cycle as a possible cause for the tendency 
of warm events to peak in boreal winter, as well as for ENSO's 
intriguing mix of temporal regularities and irregularities. 
The mechanisms by which this interaction takes place are numerous 
and intricate and their relative importance is not yet fully 
understood \citep{Tzi+95,Batt88}.

\subsection{Formulation of DDE models}
Starting in the 1980s, the effects of {\it delayed feedbacks} and 
{\it external forcing} have been studied using the formalism of 
{\it delay differential equations} (DDE) (see, {\it inter alia},
\citet{BG82,GC87} for geoscience applications, and \citet{Hale,Nuss} 
for DDE theory).
Several DDE systems have been suggested as toy models for ENSO variability.
Battisti and Hirst (\citeyear{BH89}) have considered the linear autonomous DDE 
\be
dT/dt=-\alpha\,T(t-\tau)+T,\quad \alpha>0,~\tau>0.
\label{BH}
\ee
Here, $T$ represents the sea-surface temperature (SST) averaged over the
eastern equatorial Pacific.
The first term on the right-hand side (rhs) of \eqref{BH} mimics the 
negative feedback due to the oceanic waves, while the second term reflects
Bjerknes's positive feedback.
As shown in \citep{BH89}, Eq. \eqref{BH} reproduces some of 
the main features of a fully nonlinear coupled atmosphere-ocean model 
of ENSO dynamics in the tropics \citep{Batt88,CZ}.

Suarez and Schopf (\citeyear{SS88}) and Battisti and Hirst (\citeyear{BH89}) also studied
a nonlinear version of \eqref{BH}, in which
a cubic nonlinearity is added to the rhs of the equation:
\be
dT/dt = -\alpha\,T(t-\tau)+T-T^3, 
\label{SS}
\ee
where $\quad 0<\alpha<1$ and $\tau>0$.
This system has three steady states, obtained by finding the roots of the rhs:
\[T_0=0,\quad T_{1,2}=\pm(1-\alpha)^{1/2}.\]
The so-called {\it inner} solution $T_0$ is always unstable, while
the {\it outer} solutions $T_{1,2}$ may be stable or unstable depending
on the parameters $(\alpha,\tau)$. 
If an outer steady state is unstable, the system exhibits bounded 
oscillatory dynamics; in \citep{SS88} it was shown numerically that a typical period
of such oscillatory solutions is about two times the delay $\tau$.

The delay equation idea was very successful in explaining the periodic 
nature of ENSO events.
Indeed, the delayed negative feedback does not let a solution
fade away or blow up, as in the ordinary differential equation (ODE) case 
with $\tau=0$, and 
thus creates an internal oscillator with period depending on the delay 
and particular form of the equation's rhs.
DDE modeling has also emphasized the importance of nonlinear interactions
in shaping the complex dynamics of the ENSO cycle.
At the same time, many important details of ENSO variability still
had to be explained.
 
First, a delayed oscillator similar to \eqref{BH} or \eqref{SS} typically
has periodic solutions with well-defined periods.
However, the occurrence of ENSO events is irregular and can only be
approximated very coarsely by a periodic function.
Second, El-Ni\~no events always peak during the Northern Hemisphere
(boreal) winter, hence their name; such phase locking cannot
be explained by a purely internal delayed oscillator.
Third, the value of the period produced by the delay equations 
deviates significantly from actual ENSO interevent times of 2--7 years.
The delay $\tau$, which is the sum of the basin-transit times of the westward 
Rossby and eastward Kelvin waves, can be roughly estimated to lie in the 
range of 6--8 months.
Accordingly, model \eqref{SS} suggests a period of 1.5--2 years, at most, for
the repeating warm events; this is about half the dominant ENSO recurrence time.

Tziperman {\it et al.} (\citeyear{Tzi+94}) have demonstrated that these discrepancies
can be removed --- still within the DDE framework --- 
by considering nonlinear interactions between the internal oscillator
and the external periodic forcing by the seasonal cycle. 
These authors also introduced a more realistic type of nonlinear coupling
between atmosphere and ocean to reflect the fact that the delayed
negative feedback saturates as the absolute value of 
the key dependent variable $T$ increases; note that in \eqref{BH} the 
feedback is linearly proportional to the delayed state variable $T(t-\tau)$.  
Munnich {\it et al.} (\citeyear{MCZ91}) studied an iterated-map
model of ENSO and made a detailed comparison between cubic and sigmoid
nonlinearities. 
As a result, the sigmoid type of nonlinearity was chosen in \citep{Tzi+94},
resulting in the periodically forced, nonlinear DDE
\begin{eqnarray}
\quad\quad dT/dt&=&-\alpha\,\tanh\left[\kappa\,T(t-\tau_1)\right]\nonumber\\
&+&\beta\,\tanh\left[\kappa\,T(t-\tau_2)\right]+\gamma\,\cos(2\,\pi\,t).
\label{Tzi}
\end{eqnarray}

Model \eqref{Tzi} was shown to have solutions that possess an integer period,
are quasiperiodic, or exhibit chaotic behavior, depending on the parameter values.
The increase of solution complexity --- from period one, to
integer but higher period, and on --- to quasiperiodicity and chaos 
--- is caused by the increase of the atmosphere-ocean coupling 
parameter $\kappa$.
The study \citep{Tzi+94} also demonstrated that this forced DDE system
exhibits period locking, when the external, ``explicit'' oscillator wins
the competition with the internal, delayed one, causing the system to 
stick to an integer period; see also the more detailed analysis of
phase locking in the intermediate coupled model (ICM) of Jin {\it et al.}
(\citeyear{JNG94,JNG96}).

To summarize our motivation for the choice of a ``toy model,'' 
work during the past 30 years has shown that ENSO dynamics is governed, 
by and large, by the interplay of several nonlinear mechanisms that can be 
studied in simple forced DDE models.
Such models provide a convenient paradigm for explaining 
interannual ENSO variability and shed new light on its dynamical
properties.
So far, though, DDE model studies of ENSO have 
been limited to linear stability analysis of steady-state solutions, which are
not typical in forced systems, case studies of particular trajectories, or
one-dimensional scenarios of transition to chaos, varying a single parameter 
while the others are kept fixed.
A major obstacle for the complete bifurcation 
and sensitivity analysis of such DDE models lies in the complex nature 
of DDEs, whose numerical and analytical treatment is much harder than that of 
their ODE counterparts. 

In this work we take several steps toward a comprehensive
analysis of DDE models relevant for ENSO phenomenology.
In doing so, we also wish to illustrate the complexity of phase-parameter
space structure for even such a simple model of climate dynamics.

In Section \ref{model}, we formulate our DDE model, provide basic theoretical
results for this type of DDEs, present the numerical integration method used,
and describe several solution types and their possible physical interpretation.
In Section \ref{CT}, we proceed to explore fully solution behavior over a broad
range of the model's three most physically relevant parameters.
We reproduce several dynamical solution features and bifurcation scenarios 
previously reported in the literature for both simpler \citep{SG01} and 
more detailed \citep{JNG94,JNG96,Tzi+94,Tzi+95,GR00,Neel+94,Neel+98,Dij06} models, 
report new ones, and describe the corresponding three-dimensional (3-D) regime diagram.
This 3-D regime diagram includes large regions of very smooth parameter
dependence, as well as regions of very sensitive dependence on the parameters;
the neutral surface separating simpler from more complex behavior exhibits
rich and apparently fractal patterns.

The discussion in Section \ref{discussion} highlights the possibility of 
spontaneous, intrinsic transitions between the presence or absence of
intraseasonal, higher-frequency fluctuations, as well as of interdecadal,
lower-frequency variability. 
Such higher- and lower-frequency variability accompanies the seasonal 
and interannual oscillations that dominate our model solutions.
This coexistence of variabilities on several time scales affects not
only the mean and period of the solutions, but also the distribution
of extreme warm and cold events. 

An illustration of the complexity ``burst'' caused by introducing a 
scalar delay in a simple ODE is given in Appendix~\ref{burst}.
Appendix~\ref{proof} contains the proof of a key
theoretical result presented in Sect.~\ref{model}, 
while Appendix~\ref{solver} provides details on numerical procedures for DDEs.

\section{Model and numerical integration method}
\label{model}

\subsection{Model formulation and parameters}
We consider a nonlinear DDE with additive, periodic forcing,
\begin{equation}
\frac{dh(t)}{dt}=-a\,\tanh\left[\kappa\,h(t-\tau)\right]+b\,\cos(2\pi\,\omega\,t);
\label{dde}
\end{equation}
here $~t\ge 0$ and the parameters $a,\kappa,\tau,b,$ and $\omega$ are all real and positive.
Equation~\eqref{dde} is a simplified one-delay version of 
the two-delay model considered by Tziperman {\it et al.} (\citeyear{Tzi+94});
it mimics two mechanisms essential for ENSO variability: 
delayed negative feedback via the function $\tanh(\kappa\,z)$ and periodic external forcing.
As we shall see, these two mechanisms suffice to generate very rich behavior
that includes several important features of more detailed models and of
observational data sets.
Including the positive Bjerkness feedback \citep{Phil90,Bjer69,Neel+94,Neel+98} 
is left for future work.

The function $h(t)$ in \eqref{dde} represents the thermocline depth deviations 
from the annual mean in the Eastern Pacific; accordingly, it can also be 
roughly interpreted as the regional SST, since a deeper thermocline 
corresponds to less upwelling of cold waters, and hence higher SST, and
vice versa. 
The thermocline depth is affected by the wind-forced, eastward Kelvin and 
westward Rossby oceanic waves. 
The waves' delayed effects are modeled by the function 
$\tanh\left[\kappa\,h(t-\tau)\right]$;
the delay $\tau$ is due to the finite wave velocity and corresponds 
roughly to the combined basin-transit time of the Kelvin and Rossby waves. 
The particular form of the delayed nonlinearity plays a very important role in 
the behavior of a DDE model.
Munnich {\it et al.} (\citeyear{MCZ91}) provide a physical justification for the 
monotone, sigmoid nonlinearity we adopt.
The parameter $\kappa$, which is the linear slope of $\tanh(\kappa\,z)$
at the origin, reflects the strength of the atmosphere-ocean coupling.
The forcing term represents the seasonal cycle in the trade winds. 

The model \eqref{dde} is fully determined by its five parameters: 
feedback delay $\tau$, atmosphere-ocean coupling strength  $\kappa$, 
feedback amplitude $a$, forcing frequency $\omega$, and 
forcing amplitude $b$. 
By an appropriate rescaling of time $t$ and dependent variable $h$, we let 
$\omega=1$ and $a=1$. 
The other three parameters may vary, reflecting different physical conditions of 
ENSO evolution.
We consider here the following ranges of these parameters:
$0\le \tau \le 2$~yr, 
$0< \kappa <\infty$,
$0\le b < \infty$.

To completely specify the DDE model \eqref{dde} we need to prescribe
some initial ``history,'' {\it i.e.} the behavior of $h(t)$ on the 
interval $[-\tau,\,0)$ \citep{Hale}.
In most of the numerical experiments below we assume $h(t)\equiv 1$, 
$-\tau\le t<0$, {\it i.e.} we start with a warm year.
Numerical experiments with alternative specifications of the initial 
history suggest that this choice does not affect our qualitative conclusions. 

\subsection{Basic theoretical results}
\label{basic}
To develop some intuition about the dynamics of Eq.~\eqref{dde},
we consider two limiting cases.
In the absence of the feedback, $a = 0$, the model becomes a simple ODE 
and hence has only a sinusoidal solution with period $1$. 
One expects to observe the same behavior for $b/a \gg 1$. 
In the absence of forcing, $b = 0$, we obtain a well-studied DDE
\be
\dot h(t)=-g\left[h(t-\tau)\right],\quad g(z)=\tanh(\kappa\,z).
\label{b0}
\ee 

The character of the solutions of this equation depends strongly on the 
delay $\tau$.
For small delays, one expects to see behavior reminiscent of
the corresponding ODE with zero delay; the general validity of such 
``small-delay expectations'' is analyzed in detail in \citep{Bod04}. 
For larger delays the nonlinear feedback might produce more complex dynamics. 
These heuristic intuitions happen to be true: 
Eq.~\eqref{b0} has an asymptotic solution that is identically zero for $\tau\le\tau_0$,
and admits periodic solutions with period $4\tau$ for $\tau> \tau_0$, 
where the critical delay is $\tau_0 = \pi/(2\,\kappa)$ \citep{Cao,Nus79,CW88}. 
In addition, it is known that the null solution is the only stable solution 
for $\tau\le\tau_0$.
At $\tau=\tau_0$ the system undergoes a Hopf bifurcation, and the trivial
steady state transfers its stability to a periodic solution.
Among other solutions, an important role is played by the so-called
{\it slow oscillating solutions}, whose zeros are separated by
a distance of at least the delay $\tau$.
In particular, Chow and Walther (\citeyear{CW88}) showed that periodic solutions
with period $4\,\tau$ and the symmetry condition $-h(t)=h(t-2\,\tau)$ are
exponentially asymptotically stable; that is, any other solution will
approach one of these solutions at an exponential rate.
Moreover, for $\tau>\tau_0$, these solutions
may be the only stable ones \citep{CW88}.

These results can help explain the observed near-periodicity of ENSO variability.
If one takes the delay $\tau$ to equal approximately the transit time of the 
traveling ocean waves, namely 6 to 8 months, then the $4\,\tau$ internal period 
of the ENSO oscillator becomes 2--3 years.
This remark, together with our further observations in Sect.~\ref{examples},
provides a good justification for the observed quasi-biennial oscillation
in Tropical Pacific SSTs and trade winds \citep{Diaz92,Phil90,Jiang+95,Ghil+02}.   

Below we summarize basic theoretical results about Eq.~\eqref{dde}.
Following the traditional framework \citep{Hale,Nuss}, we consider the Banach
space $X=C([-\tau,0),\mathbb{R})$ of continuous functions 
$h\,:\,[-\tau,0)\to\mathbb{R}$ and define
the norm for $h\in X$ as
\[\parallel h\parallel=\sup\left\{|h(t)|,~t\in[-\tau,0)\right\},\]
where $|\cdot|$ denotes the absolute value in $\mathbb{R}$.
For convenience, we reformulate the DDE initial-value problem (IVP)
in its rescaled form:
\begin{eqnarray}
\frac{dh(t)}{dt}&=&-\tanh\left[\kappa\,h(t-\tau)\right]+b\,\cos(2\pi\,t),~t\ge 0,\label{ivp1}\\
\quad h(t)&=&\phi(t),~t\in[-\tau,\,0),\quad\phi(t)\in X\label{ivp2}.
\end{eqnarray}

\begin{proposition}{\bf (Existence, uniqueness, continuous dependence)}
For any fixed positive triplet $(\tau,\,\kappa,\,b)$, 
the IVP \eqref{ivp1}-\eqref{ivp2} has a unique solution $h(t)$ 
on $[0,\,\infty)$.
This solution depends continuously on the initial data $\phi(t)$, 
delay $\tau$ and the rhs of \eqref{ivp1} considered as a continuous map
$f\,:\,[0,T)\times X\to \mathbb{R}$, for any finite $T$.
\label{prop}
\end{proposition}
{\bf Proof.} See Appendix~\ref{proof}.\\

From Proposition~\ref{prop} it follows, in particular, that the 
system \eqref{ivp1}-\eqref{ivp2} has a unique solution for all time, 
which depends continuously on the model parameters $(\tau,\,\kappa,\,b)$
for any finite time.
This result implies that any discontinuity in the solution profile as
a function of the model parameters indicates the existence of an
unstable solution that separates the attractor basins of two stable solutions.

Our numerical experiments suggest, furthermore, that all stable solutions 
of \eqref{ivp1}-\eqref{ivp2} are bounded and have an infinite number of zeros.

\subsection{Numerical integration}
The results in this study are based on numerical integration of the 
DDE \eqref{ivp1}-\eqref{ivp2}.
We emphasize that there are important differences between numerical
integration of DDEs and ODEs.
The first important difference is in the requisite initial data.
The solution of a system of ODEs is determined by its value
at the initial point $t = t_0$.
When integrating a DDE, terms like $h(t - \tau)$ may 
represent values of the solution at points prior to the initial point.
Because of this, the given initial data must include not only
$h(t_0)$, but also a ``history'' of the values $h(t)$ for all $t$
prior to $t_0$ in the interval that extends as far back at the largest delay.

Another important issue in solving DDEs arises when a delayed value of the argument 
falls within the current integration step.
In order to avoid limiting the step size to be smaller than
the smallest delay, or, alternatively, to avoid extrapolating the previous
solution, an iterative procedure must be used to obtain successive
approximations of the delayed solution that will yield a satisfactory
local error estimate.
The implementation of this iterative procedure
affects profoundly the performance of a DDE solver.

These and other specific features of DDE numerical integration 
require development of special software and often the problem-specific 
modification of such software.
We used here the Fortran 90/95 DDE solver \texttt{dde\_solver} of 
Shampine and Thompson (\citeyear{F90}), available at
\url{http://www.radford.edu/~thompson/ffddes/}.
This solver implements a (5,6) pair of continuously embedded, explicit 
Runge-Kutta-Sarafyan methods \citep{dklag6}.
Technical details of \texttt{dde\_solver}, as well as a brief overview 
of other available DDE solvers are given in Appendix ~\ref{solver}. 

The numerical simulations in this paper require very long
integration intervals, leading to prohibitive storage requirements.
This difficulty led us to incorporate several new options in \texttt{dde\_solver};
they are also described in Appendix~\ref{solver}.

\subsection{Examples of model dynamics}
\label{examples}
\begin{figure}[t]
\vspace*{2mm}
\centering\includegraphics[width=8.3cm]{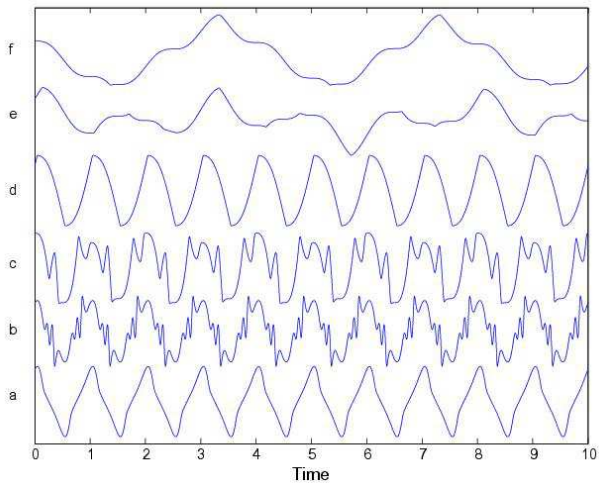} 
\caption{Examples of DDE model solutions.
Model parameters are $\kappa=100$ and $b=1$, while
$\kappa$ increases from curve (a) to curve (f) as follows:
(a) $\tau=0.01$,
(b) $\tau=0.025$,
(c) $\tau=0.15$,
(d) $\tau=0.45$,
(e) $\tau=0.995$, and
(f) $\tau=1$.} 
\label{tr_ex}
\end{figure}

In this subsection we illustrate typical solutions of the problem 
\eqref{ivp1}-\eqref{ivp2} and comment on physically relevant 
aspects of these solutions.
All experiments shown here use the constant initial data $\phi\equiv 1$.
Figure \ref{tr_ex} shows six trajectories obtained by fixing
$b=1,\kappa=100$ and varying the delay $\tau$ over two orders
of magnitude, from $\tau=10^{-2}$ to $\tau=1$, with $\tau$ increasing
from bottom to top in the figure.
The sequence of changes in solution type illustrated in Fig.~\ref{tr_ex}
is typical for any choice of $(b,\kappa)$ as $\tau$ increases.

For a small delay, $\tau<\pi/(2\,\kappa)$, we have a periodic solution with 
period 1 (curve a); here the internal oscillator is completely dominated 
by the seasonal forcing.
When the delay increases, the effect of the internal oscillator
becomes visible: small wiggles, in the form of amplitude-modulated oscillations 
with a period of $4\,\tau$, 
emerge as the trajectory crosses the zero line.
However, these wiggles do not affect the overall period, which is still unity.
The wiggle amplitude grows with $\tau$ (curve b) and eventually wins
over the seasonal oscillations, resulting in period
doubling (curve c).  
Further increase of $\tau$ results in the model passing through a sequence of
bifurcations that produce solution behavior of considerable interest for 
understanding ENSO variability.

\begin{figure}[t]
\vspace*{2mm}
\centering\includegraphics[width=8.3cm]{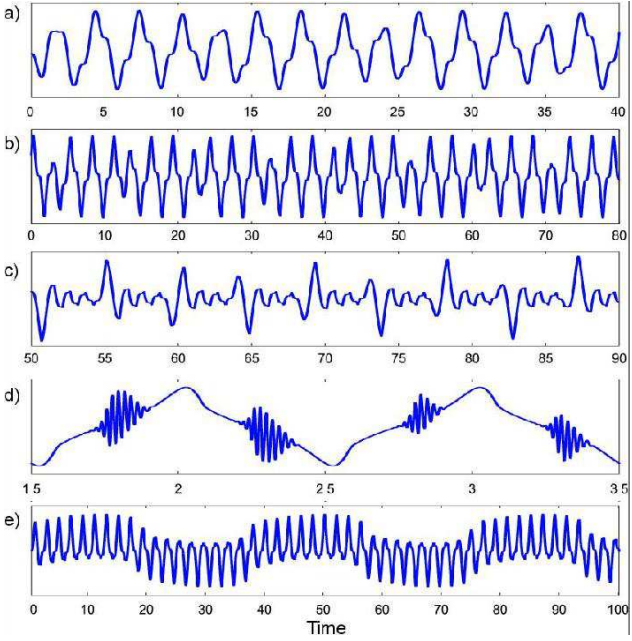}
\caption{Noteworthy solution patterns of relevance to ENSO dynamics;
seasonal forcing amplitude $b=1$.
a) Regularly occurring cold (low-$h$) events, or La Ni\~nas ($\kappa=5$, $\tau=0.65$);
b) irregular cold events ($\kappa=100$, $\tau=0.58$); 
c) irregular alternations of warm (El Ni\~no, high-$h$) and cold events ($\kappa=50$, $\tau=0.42$);
d) intraseasonal activity reminiscent of Madden-Julian oscillations
or westerly-wind bursts ($\kappa=500$, $\tau=0.005$); and
(e) interdecadal variability in the annual mean and in the relative amplitude of warm and
cold events ($\kappa=50$, $\tau=0.508$).} 
\label{nw}
\end{figure}

Some of these types of solution behavior are illustrated further 
in Fig.~\ref{nw}.
Panel (a) ($\kappa=5$, $\tau=0.65$) shows the occurrence of ``low-$h$," 
or cold, events every fourth seasonal cycle.
Note that negative values of $h$ correspond to the boreal 
(Northern Hemisphere) winter, that is to the upwelling season --- 
December-January-February --- in the eastern Tropical Pacific;
in the present, highly idealized model, we can associate the 
extreme negative values with large-amplitude cold events, or La Ni\~nas.
This solution pattern loses its regularity when the atmosphere-ocean
coupling increases:
Panel b ($\kappa=100, \tau=0.58$) shows irregular occurrence
of large cold events with the interevent time varying from 3 to 7 cycles. 

In panel c ($\kappa=50, \tau=0.42$) we observe alternately and irregularly
occurring warm El-Ni\~no and cold La Ni\~na events: 
the ``high-$h$'' events occur with a period of about 4 years and random magnitude.
Panel d ($\kappa=500, \tau=0.005$) shows another interesting type of 
behavior: bursts of intraseasonal oscillations of random amplitude 
superimposed on regular, period-one dynamics.
This pattern is reminiscent of Madden-Julian oscillations \citep{MJ71,MJ72,MJ94}
or westerly-wind bursts \citep{Geb+07,HG88,Ver98,Del+93}.
The solution in panel e ($\kappa=50, \tau=0.508$) demonstrates sustained
interdecadal variability in the absence of any external source of such
variability.
The solution pattern illustrates spontaneous changes in the long-term annual mean,
as well as in the distribution of positive and negative extremes, 
with respect to both time and amplitude.

\section{Critical transitions}
\label{CT}

\subsection{Numerical characterization of solution behavior}
\label{stat}
In this section we focus on the onset of instabilities in the model 
\eqref{ivp1}-\eqref{ivp2}.
Taking a ``metric'' approach to the problem, we study the change in several
statistics of a trajectory as the model parameters change.
This approach is complementary to the ``topological'' one, which forms
the basis for the stability analysis of dynamical systems \citep{AP37,KH}.
In the latter, one studies the topological structure of 
the system's attractor(s), {\it i.e.} a combination of points, circles, 
tori, or more complicated objects.
The motivation for this approach comes from noting that topologically equivalent 
solutions can be mapped onto each other using an appropriate diffeomorphism,
{\it i.e.} a one-to-one, continuously differentiable map. 
Hence, considering topologically equivalent classes rather than all individual
solutions is enough for studying the system's qualitative behavior.
From a practical point of view, though, metric properties of a solution might
be as important as its topological description or more: think of living in a region
with constant air temperature of --10$^{\circ}$C vs. 20$^{\circ}$C.
Furthermore, metric properties are also much easier to study, in simple
models, as well as in full, 3-D general circulation models (GCMs) and
in observational data sets.

Technically, we proceed in the following way.
For each fixed triplet of parameters $(b,\kappa,\tau)$ we find a numerical
approximation $\hat h_i\equiv \hat h(t_i)$ of the model solution $h(t)$ on a grid 
$\cG=\{t_i\}_{i=1,\dots,N}$, 
$t_i\in [0,\,T_{\rm max}]$ using the Fortran 90 DDE solver \texttt{dde\_solver} of
Shampine and Thompson (\citeyear{F90}) (see also Appendix~\ref{solver}).
We only consider the latter part of each solution, $t_i>T_{\min}>0$, in order to avoid 
any transient effects; to simplify notations, we assume from now on
that $t_1=T_{\rm min}$.
Typical parameters for our numerical experiments are $T_{\rm max}=10^4$,
$T_{\rm max}-T_{\rm min}=10^3$, time step $\delta=t_i-t_{i-1}=10^{-3}$, and a
numerical precision of $\epsilon=10^{-4}$.
We have also verified some of the results with a precision up to $10^{-12}$,
a time step of $10^{-4}$, and over time intervals up to $T_{\rm max}-T_{\rm min}=10^4$
in order to ascertain that the reported phenomena are not caused or affected 
by numerical errors.  

We report results for the following trajectory statistics:
maximum value $M=\max_{i}\,\hat h_i$;
mean value $E=\sum_i \hat h_i/N$, where $N$ is the maximal integer less
than $(T_{\rm max}-T_{\rm min})/\delta$; and
mean of positive values $E_+=\sum_i \hat h_i{\bf 1}_{\hat h_i>0}/\sum_i {\bf 1}_{\hat h_i>0}$,
where ${\bf 1}_A$ is the characteristic function of the set $A$,
identically equal to one for all points in $A$ and zero outside of $A$.
Furthermore, we have computed, but do not show here, the trajectory variance, 
the mean of negative values, and upper 90\% and 95\% quantiles; the results are 
very similar for all the statistics we have examined.

We also computed an approximation to the period of a solution.
Specifically, for any positive integer $\Delta$ we define the $\Delta$-discrepancy 
\begin{eqnarray}
R_{\Delta}&=&\frac{\sum_{i=\Delta+1}^N \left(\hat h_i-\hat h_{i-\Delta}\right)^2}
{(N-\Delta)\,Var(\hat h)},\\\
Var(\hat h)&=&\frac{1}{N}\sum_{i=1}^N\,\left(\hat h_i-\bar h\right)^2,
\quad
\bar h=\frac{1}{N}\sum_{i=1}^N \hat h_i.\nonumber
\end{eqnarray}
If $\hat h_i\equiv h(t_i)$, for a periodic solution $h$ with period $P=\delta\,\Delta$, 
we have $R_{\Delta}=0$. 
In numerical experiments, we can only guarantee that the $\Delta$-discrepancy of
a periodic solution is small enough: $R_{\Delta}<r_{\epsilon}=4\,\epsilon^2/Var(\hat h)$,
where $\epsilon$ is the absolute numerical precision.
We call {\it near-period} the minimal number $\PP=\delta\,\Delta$ such that
\[R_{\Delta}<r_{\eta}~{\rm and~} R_{\Delta}<R_{\Delta\pm 1}\]
for some prescribed $0< \eta \ll 1$.
The first condition ensures that the $\Delta$-discrepancy $R_{\Delta}$ is small enough,
while the second one guarantees that $R_{\Delta}$ is a local minimum as a function
of $\Delta$.
The following proposition follows readily from the definition of $\PP$.

\begin{proposition} {\bf (Convergence theorem).}
The near-period $\PP$ converges to the actual period $T$ 
of a continuous periodic function $h(t)$ when the numerical 
step $\delta$ and nominal accuracy $\epsilon$ decrease:
$\lim_{\delta\to 0,\, \epsilon\to 0}\PP=T.$
\label{lim}
\end{proposition}

The continuity requirement in the hypothesis is too restrictive for this statement, 
but it suffices for our purpose, since solutions of model \eqref{ivp1}, for a 
given triplet $(b,\,\tau,\,\kappa)$, are smooth.
The rate of convergence in Proposition \ref{lim} depends strongly on the period
structure of $h(t)$. 
The rate is high for functions with ``simple'' periods, {\it e.g.} with a single
local maximum within the period $T$,
and may be arbitrarily low in the general case; {\it e.g.}
for $h=h_1+\gamma\,h_2$, where $h_1$ has period $T/2$, $h_2$ period $T$,
and $\gamma$ is small enough, the convergence is quite slow.

\begin{figure}[t]
\vspace*{2mm}
\centering\includegraphics[width=8.3cm]{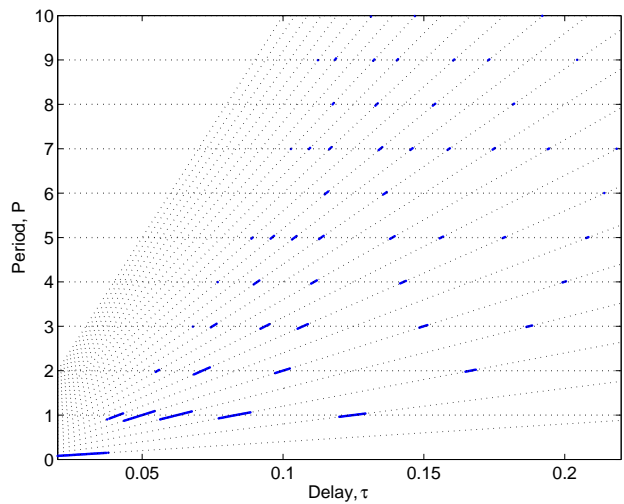}
\caption{Near-period $\PP$ as a function of delay $\tau$ for fixed
$b=0.03$, heavy solid segments; 
dashed straight lines correspond to $\PP=4\,\tau\,k$,
$k=1,2,...$.
The near-period is always a multiple of $4\,\tau$ close to an integer.} 
\label{fig2}
\end{figure}

To summarize, the near-period $\PP$ approximates the actual 
period $T$ for {\it periodic} functions.
The functional $\PP$ can also be defined for certain functions that are not
periodic.
For instance, this may be the case for a quasiperiodic function
$h(t)=p_1(t)+p_2(t),$
where each $p_i(t)$ is periodic and the two periods, $T_1$ and $T_2$, are 
rationally independent.
It is also the case for a near-periodic function
$h(t)=p_0(t)+p_1(t),$
where $p_0(t)$ is periodic, and $p_1(t)$ has sufficiently small amplitude.
As we shall see, the period approximation $\PP$ is quite helpful in understanding the 
structure of our model's solution set.

\subsection{Small forcing amplitude and frequency locking}
We mentioned in Sect.~\ref{basic} that, without external forcing ($b=0$), the
nontrivial stable solutions of the model \eqref{ivp1}-\eqref{ivp2} are periodic with
period $4\,\tau$. 
When the external forcing is small, $b\ll 1$, the dynamical system tries
to retain this property. 
Figure \ref{fig2} shows the near-period $\PP$ as a function of the delay
$\tau$ for fixed $b=0.03$ and $\kappa=100$.
Here $r_{\eta}=R_1$, that is we compare the $\Delta$-discrepancy $R_{\Delta}$ 
with the one-step discrepancy $R_1$; the latter measures the degree
of continuity of our discrete-time approximation to $h(t)$.
Straight dashed lines in the figure correspond to $\PP=4\,\tau\,k$ 
for positive integer $k$.
One can see that the solution's near-period is always a multiple of $4\,\tau$
and always close to an integer.
This state of affairs is a natural compromise between the internal period 
$4\,\tau$ and the driving period 1, a compromise rendered possible by
the internal oscillator's nonlinearity.

\begin{figure}[t]
\vspace*{2mm}
\centering\includegraphics[width=8.3cm]{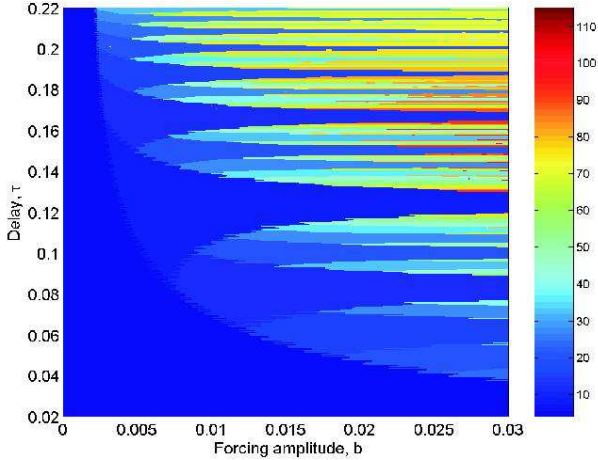}
\caption{Devil's bleachers: period index $k=\PP/(4\,\tau)$ as a function of 
forcing amplitude $b$ and delay $\tau$. 
Notice the presence of very long periods, of 100 years and more, in
the color bar and figure.} 
\label{fig1}
\end{figure}

More generally, Fig.~\ref{fig1} shows the map of 
the {\it period index} $k=\PP/(4\,\tau)$ in the 
$b$--$\tau$ plane for $0<b<0.03$, $0<\tau<0.22$.
Here one immediately recognizes the so-called Devil's bleachers
scenario of transition to chaos, documented in other ENSO models,
including both ICMs \citep{JNG94,JNG96,Tzi+94,Tzi+95,GR00} 
and GCMs \citep{GR00}, as well as in certain observations \citep{GR00,YL94}.
The periodically forced model \eqref{ivp1} exhibits the web of resonances that
characterizes coupled oscillators, although the ``external oscillator'' 
quickly wins over the internal one as the amplitude $b$ of the forcing increases.
For $b=0.1$ (not shown) the near-period plot looks similar to the
one shown in Fig.~\ref{fig2}, but $\PP$ takes only integer values.
When the forcing amplitude increases further, the near-period $\PP$, as well
as the actual period $T$, is locked solely to integer values.

\subsection{Onset of the instabilities}
Tziperman {\it et al.} (\citeyear{Tzi+94}) reported that the onset
of chaotic behavior in their two-delay, periodically forced DDE 
model is associated with the
increase of the atmosphere-ocean coupling $\kappa$;
Munnich {\it et al.} (\citeyear{MCZ91}) made a similar observation
for an iterated-map model of ENSO.
We explore this transition to chaos in our model over its entire, 
3-D parameter space. 

\begin{figure}[t]
\vspace*{2mm}
\centering\includegraphics[width=8.3cm]{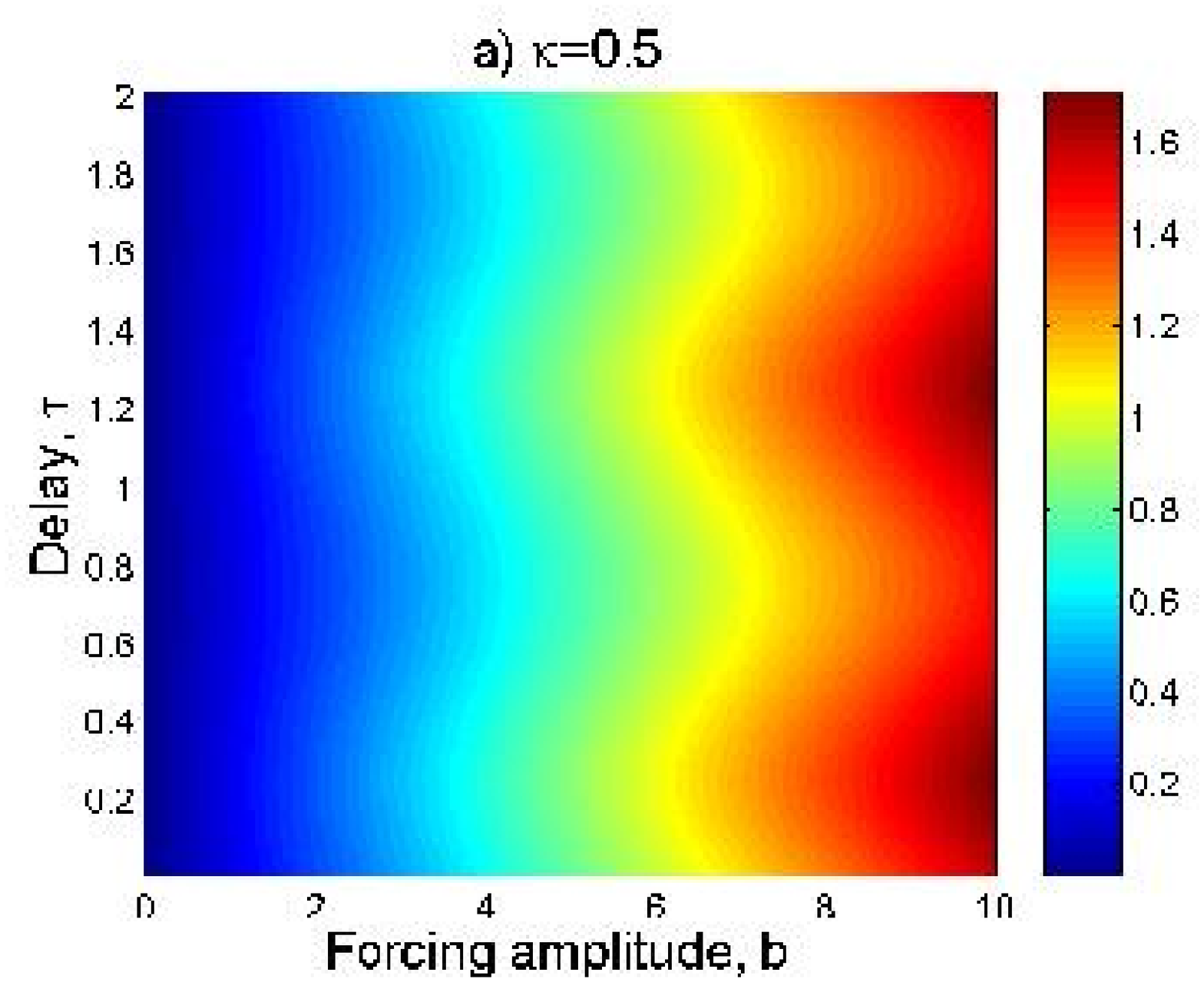}
\centering\includegraphics[width=8.3cm]{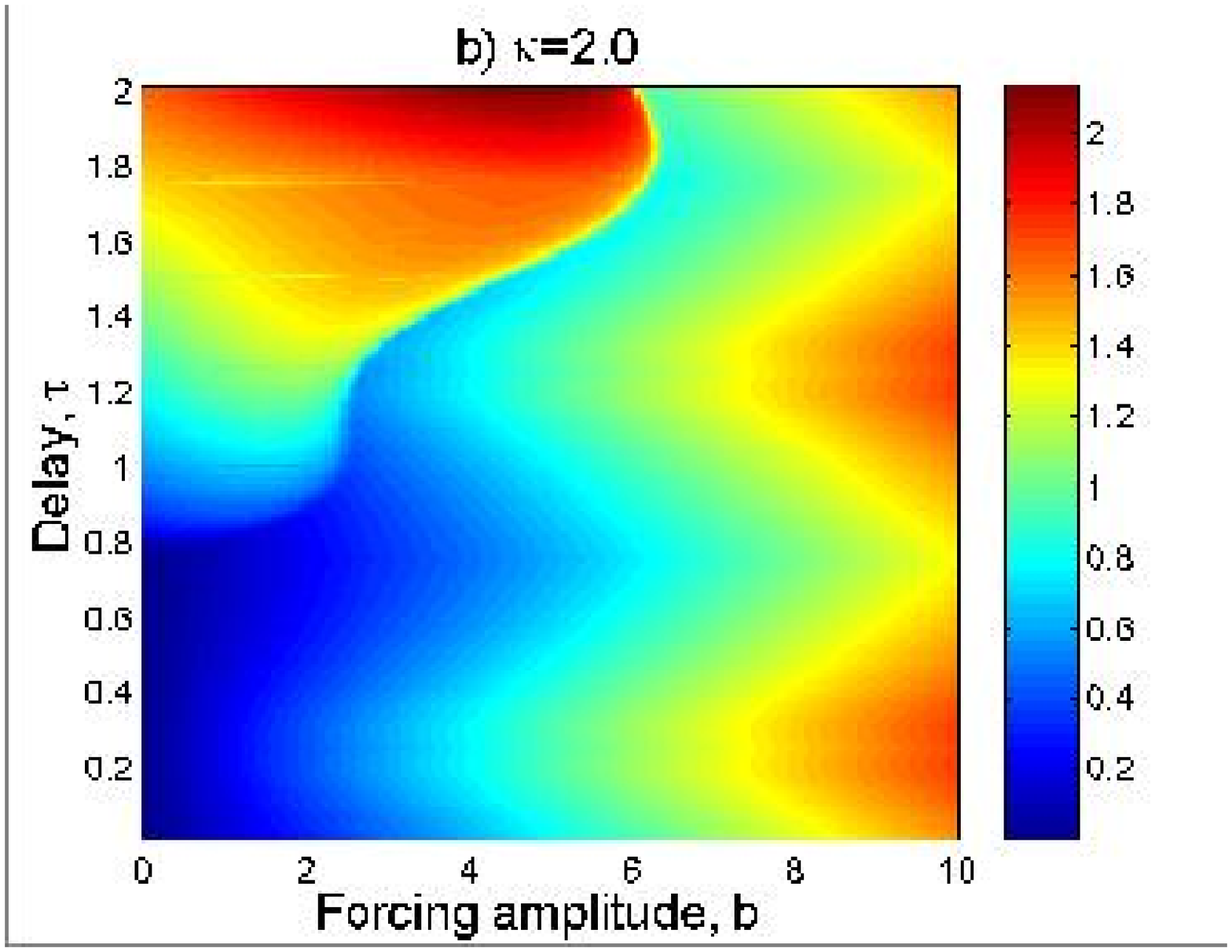}
\centering\includegraphics[width=8.3cm]{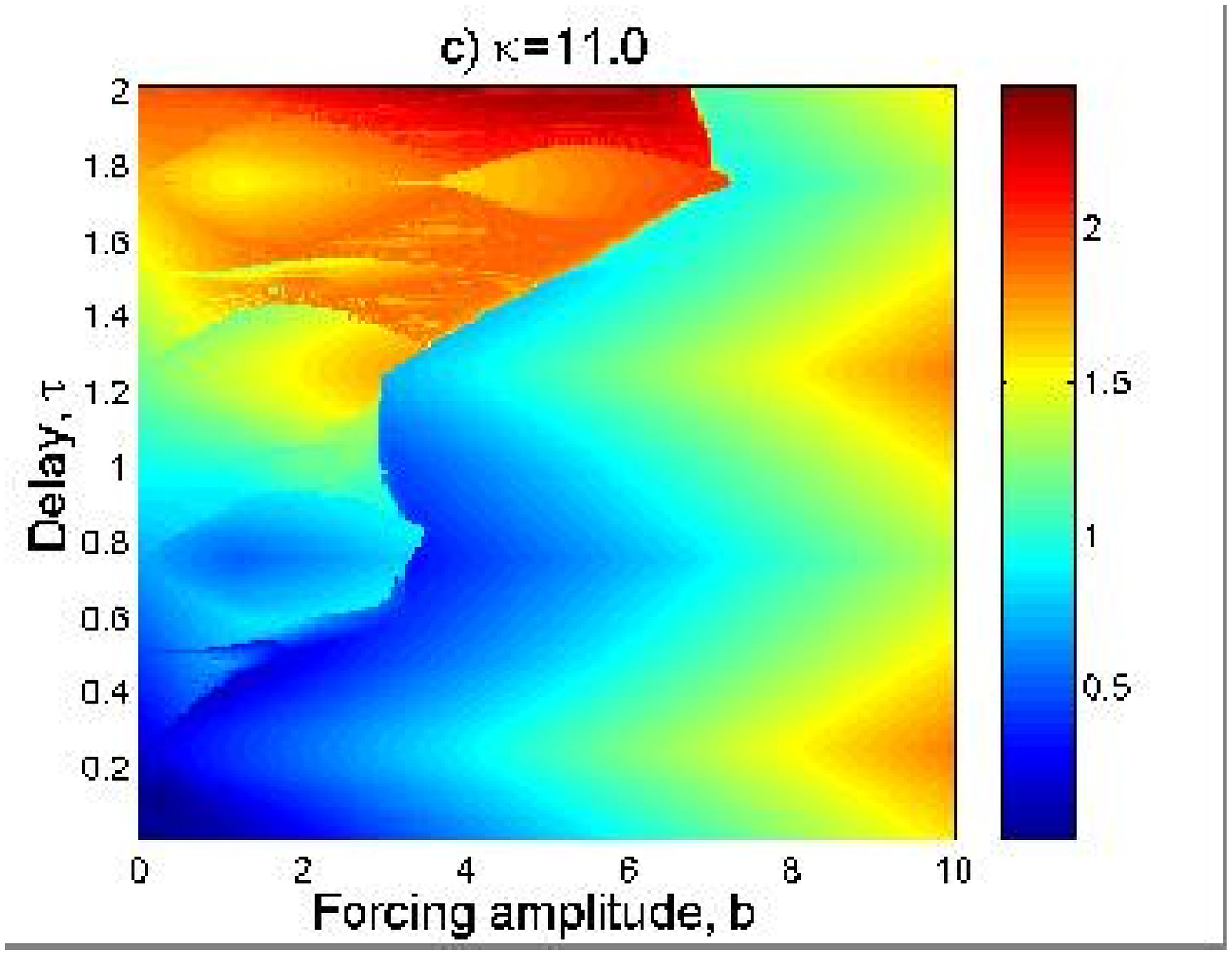}
\caption{Maximum map $M=M(b,\,\tau)$. Top: $\kappa=0.5$,
middle: $\kappa=2$, and bottom: $\kappa=11$.
Notice the onset of instabilities and emergence of a neutral curve
$f(b,\,\tau)=0$ that separates the smooth from the unstable regions.} 
\label{mmap}
\end{figure}

First, we compute in Fig.~\ref{mmap} the trajectory maximum $M$ as a function of 
the parameters $b$ and $\tau$ for increasing values of $\kappa$.
For small values of coupling (top panel) we have a smooth map, monotonously
increasing in $b$ and periodic with period 1 in $\tau$.
As the coupling increases, the map loses its monotonicity in $b$
and periodicity in $\tau$ for large values of $\tau$, but it is still smooth.
For $\kappa\approx 2$ (middle panel), a neutral curve $f(b,\tau)=0$
emerges that separates a smooth region (to the right of the curve),
where we still observe monotonicity in $b$ and periodicity in $\tau$,
from a region with rough behavior of $M$.
The gradient of $M(b,\tau)$ is quite sharp across this neutral curve.
 
Further increase of the coupling results in a qualitative change in
the maximum map.
The neutral curve, which becomes sharp and rough, separates two 
regions with very different behavior of $M(b,\tau)$ (bottom panel).
To the right of the curve, the map $M(b,\tau)$ is still smooth,
periodic in $\tau$ and monotonic in $b$.
To the left, one sees discontinuities that produce rough and 
complicated patterns.
The mean position of the neutral curve $f(b,\tau)=0$ quickly converges 
to a fixed profile, although its detailed shape at smaller scales
continues to change with increasing $\kappa$.
The limiting profile is close to the one observed for $\kappa=11$ 
(bottom panel).

\subsection{Unstable behavior}
In this subsection we illustrate the model's parametric instabilities using 
the four trajectory statistics introduced in Sect.~\ref{stat}: maximum $M$, mean $E$, 
mean of positive values $E_+$, and near-period $\PP$.
Figure~\ref{stat_map_a} shows a plot of these statistics in a rectangle of 
the plane $(b,\,\tau)$ for fixed $\kappa=10$.
The neutral curve of Fig.~\ref{mmap}c crosses this rectangle from its bottom 
left corner to the central point on its right edge; thus the bottom right region of each
panel corresponds to smooth behavior of each statistic map, while
the top left region corresponds to rough behavior. 
This figure illustrates the following points:

\begin{figure*}[t]
\vspace*{2mm}
\centering\includegraphics[width=8.3cm]{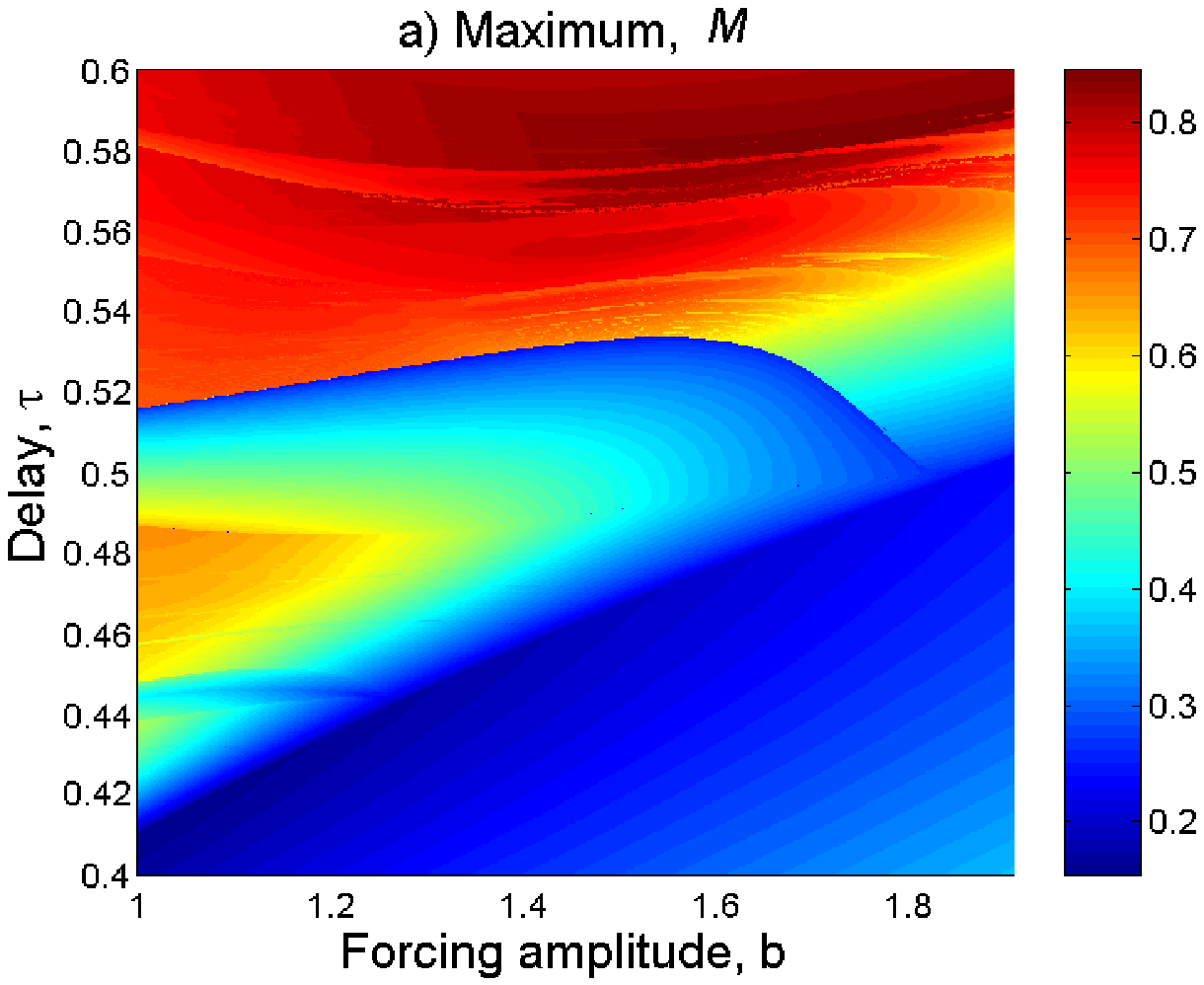}
\centering\includegraphics[width=8.3cm]{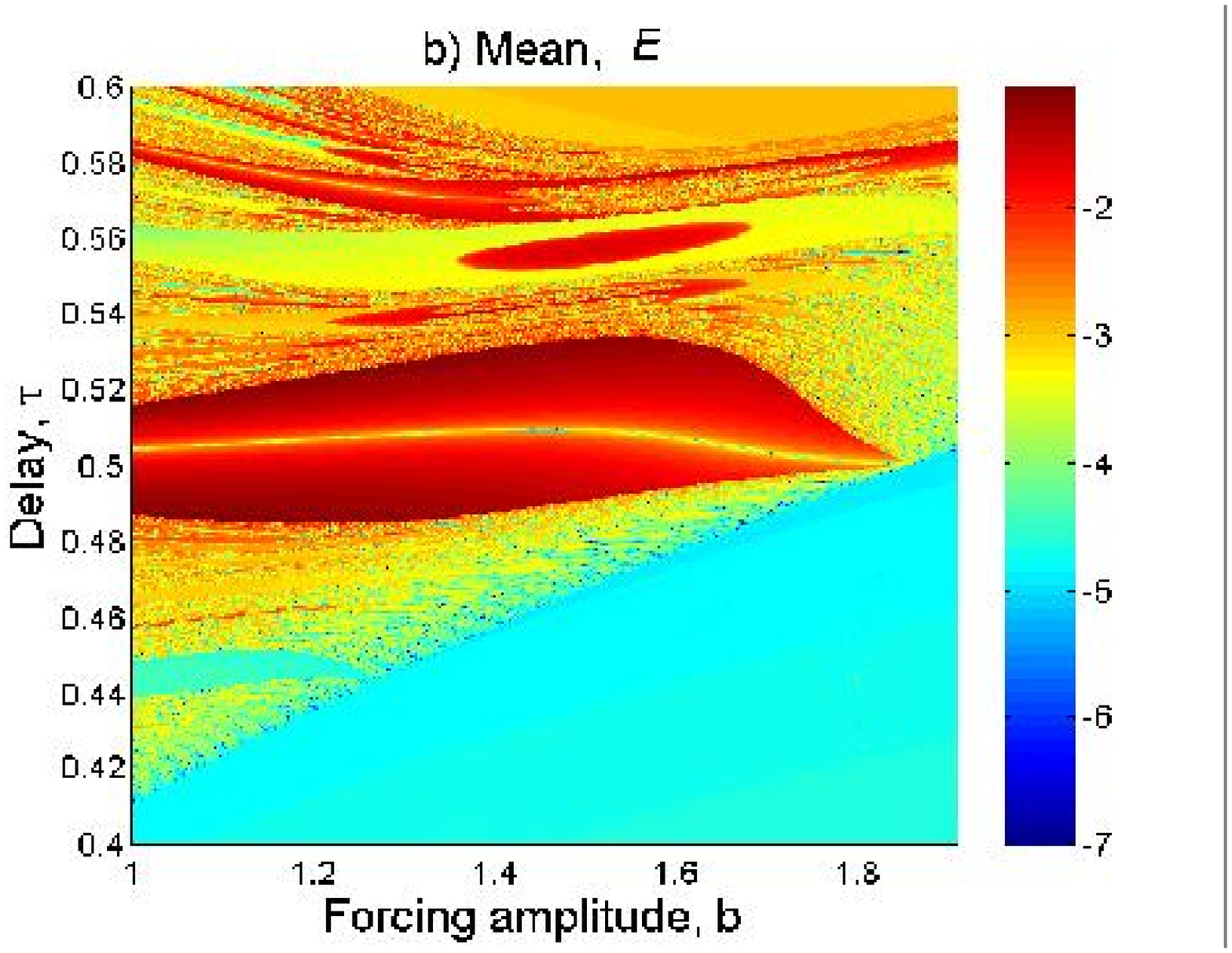}

\vspace{.5cm}
\centering\includegraphics[width=8.3cm]{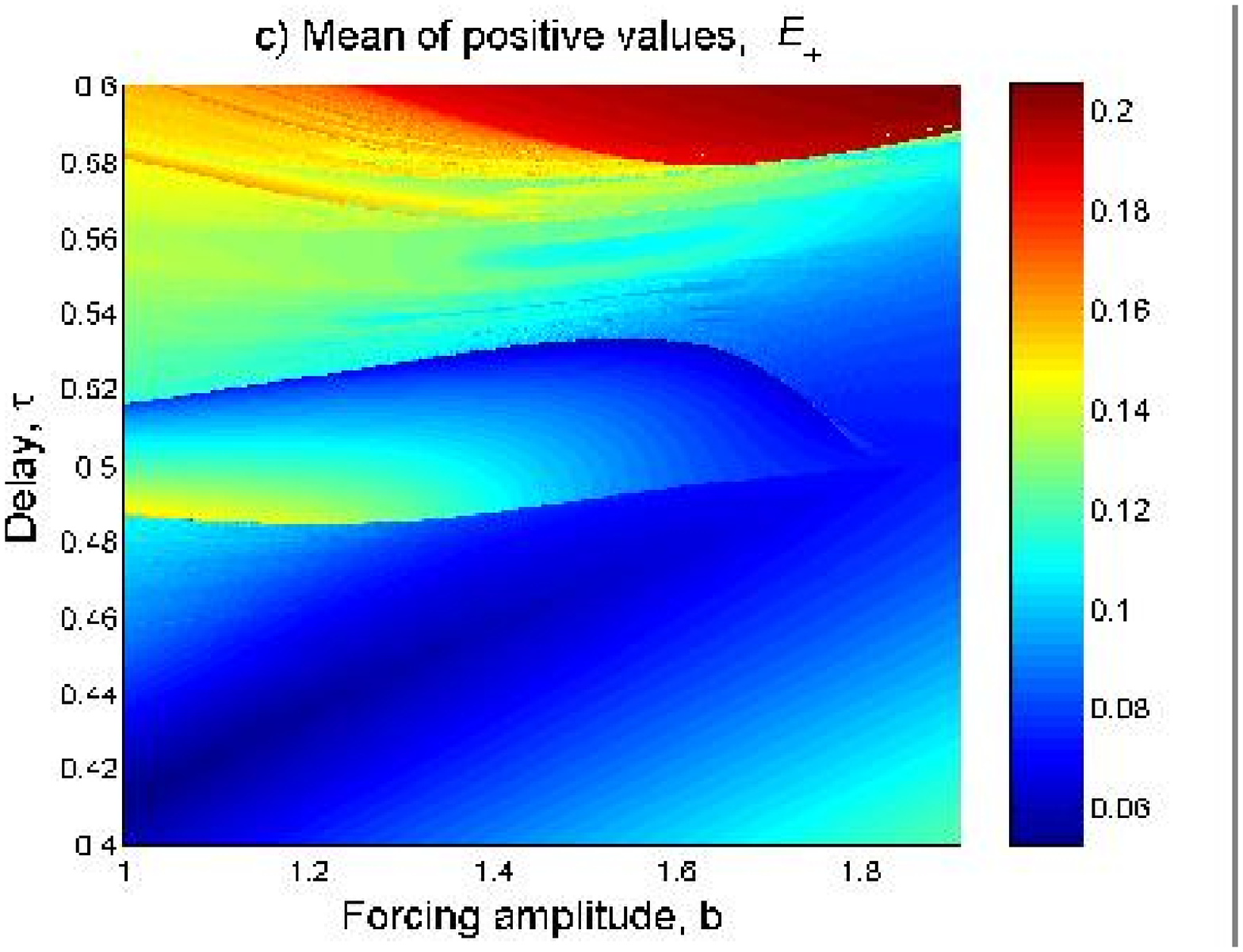}
\centering\includegraphics[width=8.3cm]{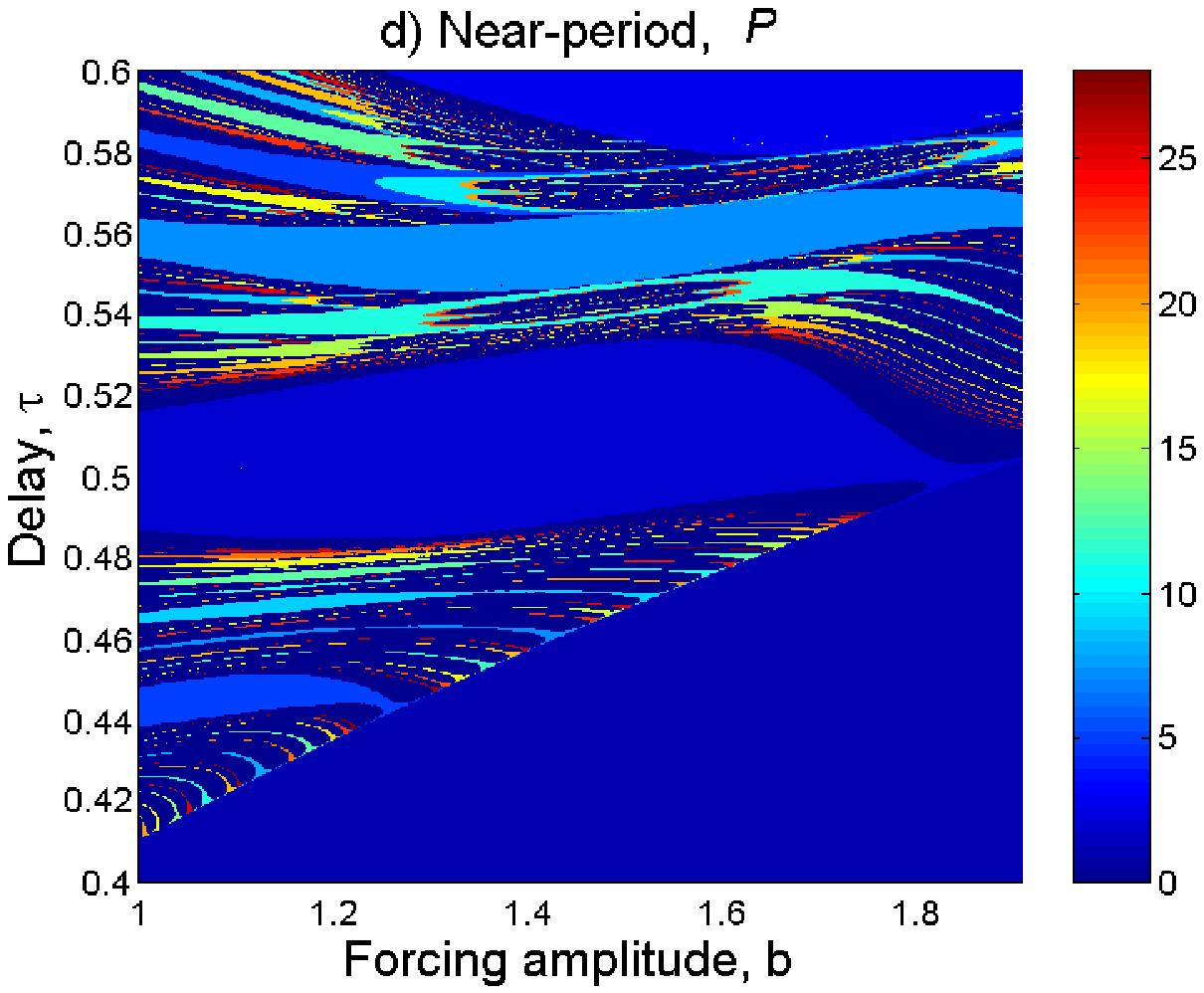}
\caption{Trajectory statistics plots for $\kappa=10$, as
a function of forcing amplitude $b$ and delay $\tau$:
a) maximum map, $M(b,\,\tau)$;
b) mean map, $\log |E(b,\,\tau)|$;
c) mean of positive values, $E_+(b,\,\tau)$; and
d) near-period map, $\PP(b,\,\tau)$.} 
\label{stat_map_a}
\end{figure*}

{\it The maximum map} $M(b,\tau)$ (Fig.~\ref{stat_map_a}a) shows, 
among other instabilities, a pronounced jump along a mainly horizontal 
curve in the $(b,\,\tau)$ plane.
In the vicinity of this curve, an arbitrarily small increase in 
$\tau$ causes a 200--300\% jump (from 0.25 to 0.5--0.8) in $M$.
A pair of trajectories on either side of this transition is shown below
in Fig.~\ref{fig_instab}.

{\it The mean map} $E(b,\,\tau)$ (Fig.~\ref{stat_map_a}b). 
To emphasize the parametric instabilities, we show here $\log_{10} |E|$.
Deviations of $E$ from 0 reflect the trajectory's asymmetry;
hence the larger values of this map indicate asymmetric solutions.
In this experiment, we use a numerical precision of $10^{-4}$,
so that values of $\log_{10} |E| < -4$ effectively correspond to symmetric 
trajectories, $E=0$.
One can see that the symmetry, characteristic for trajectories from the
smooth region (bottom right part), breaks across the neutral curve.
In the unstable region, the magnitude of the asymmetry is very intermittent;
it ranges over three orders of magnitude, taking its maximal value 
in the region that corresponds to the jump in the trajectory maximum,
cf. panel (a).

{\it The mean of positive values} $E_+(b,\,\tau)$ (Fig.~\ref{stat_map_a}c), 
by comparison with the maximum map in panel (a), shows that certain internal 
instabilities may affect trajectory shape without affecting the behavior of extremes.
For instance, the maximum map $M(b,\,\tau)$ is smooth within the neighborhood 
of the point $(b=1.5,\,\tau=0.49)$, although the map $E_+(b,\,\tau)$
exhibits a discontinuity across this neighborhood. 
In fact, one arrives at the same conclusion by comparing the maximum map
to the mean map, cf. panels (a) and (b), respectively.

{\it The near-period map} $\PP(b,\,\tau)$ (Fig.~\ref{stat_map_a}d).
The near-period is varying over the interval $[0,\,27]$ in this map.
As we have noticed, not all of these values correspond to trajectories that are
actually periodic, rather than just nearly so (see Sect.~\ref{stat}) .
The large constant regions, though, do reflect the actual periods;
as a rule, they correspond to small values of $\PP$.
Examples include: 
$\PP=1$, within the smooth part of the map (bottom right); 
$\PP=2$ within the middle horizontal tongue;
$\PP=3$ within the top right part; and
$\PP=5$ in a small tongue that touches the left margin of the plot 
at $(b=1,\,\tau=0.44)$.

\begin{figure*}[t]
\vspace*{2mm}
\centering\includegraphics[width=8.3cm]{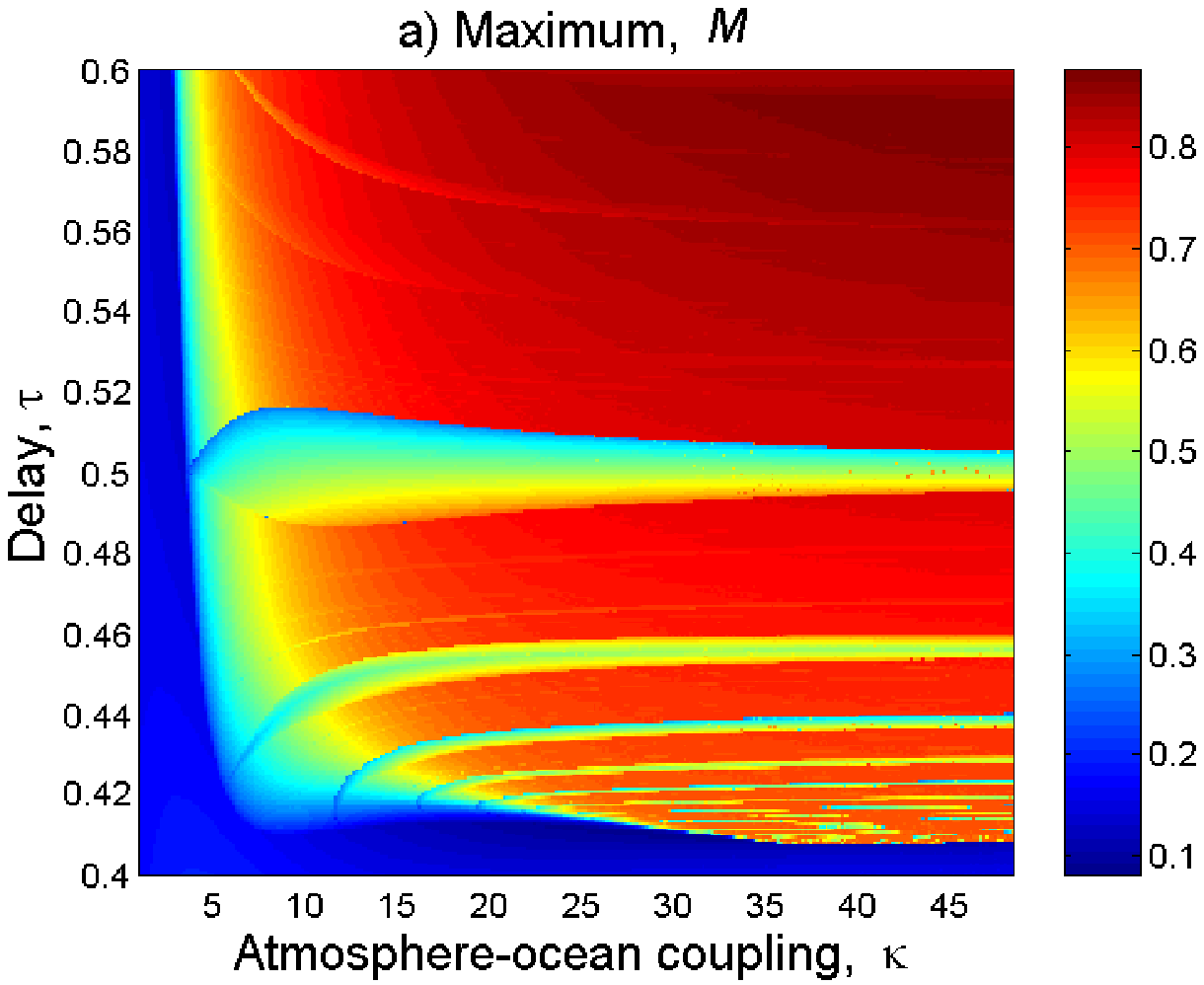}
\centering\includegraphics[width=8.3cm]{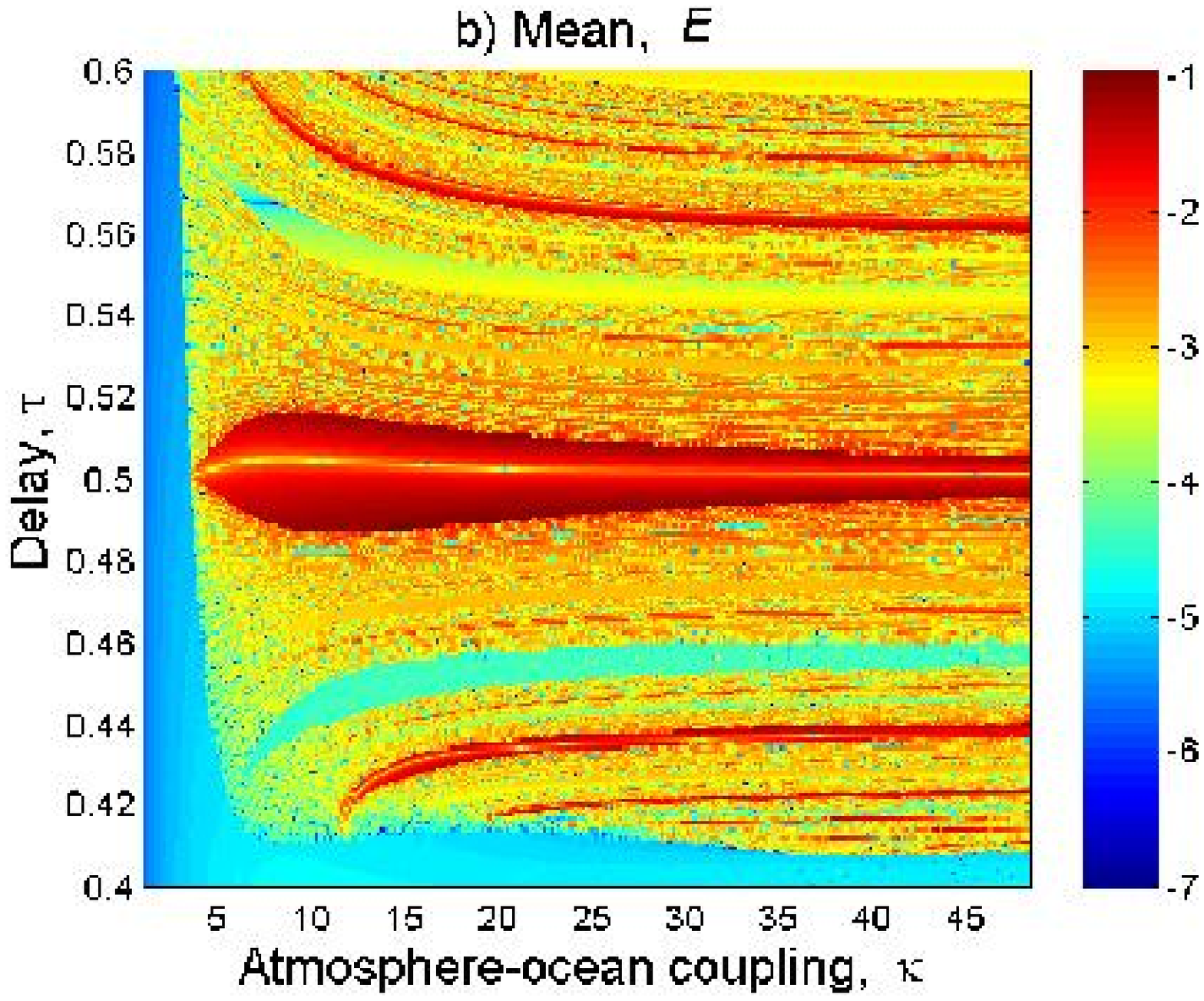}

\vspace{.5cm}
\centering\includegraphics[width=8.3cm]{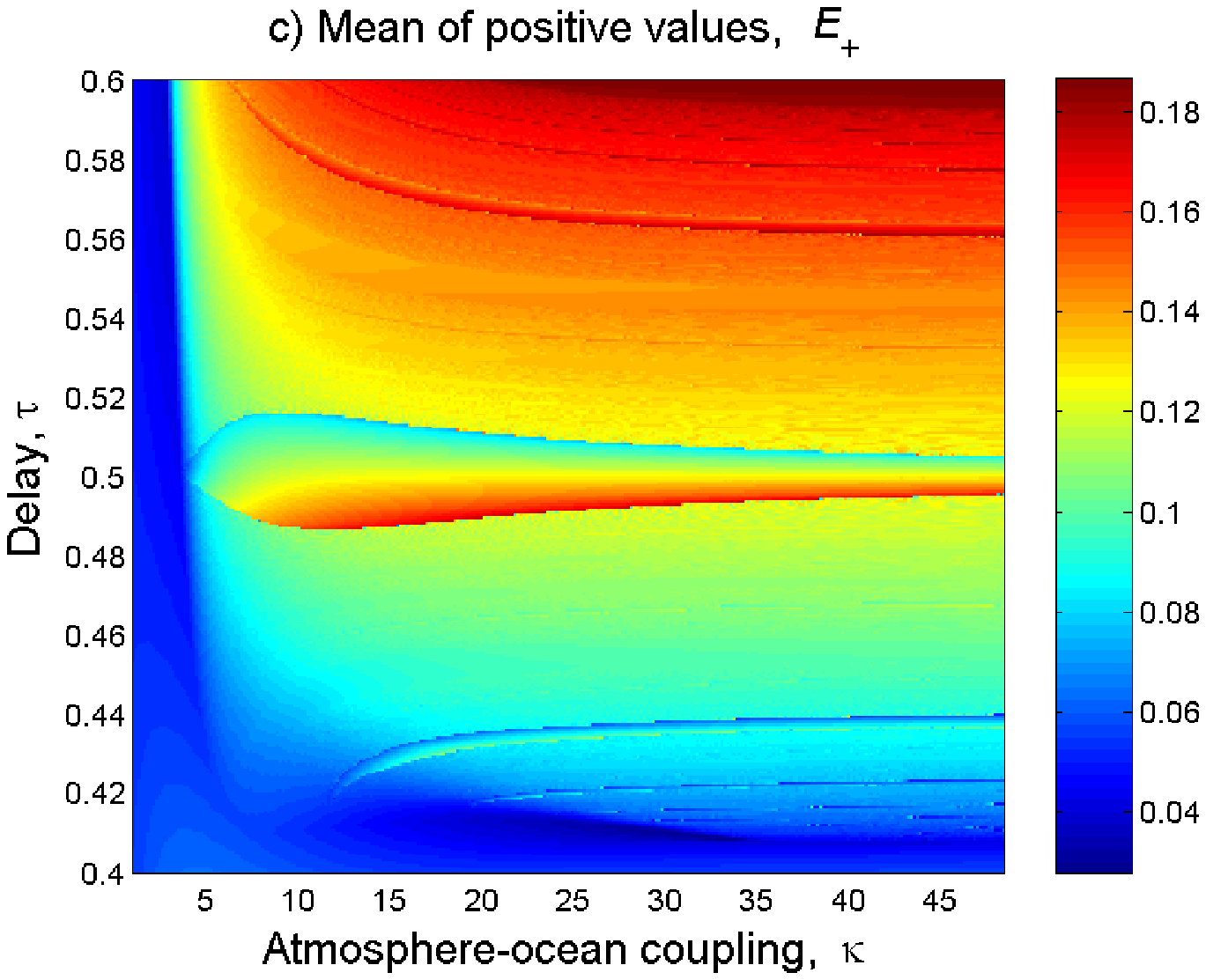}
\centering\includegraphics[width=8.3cm]{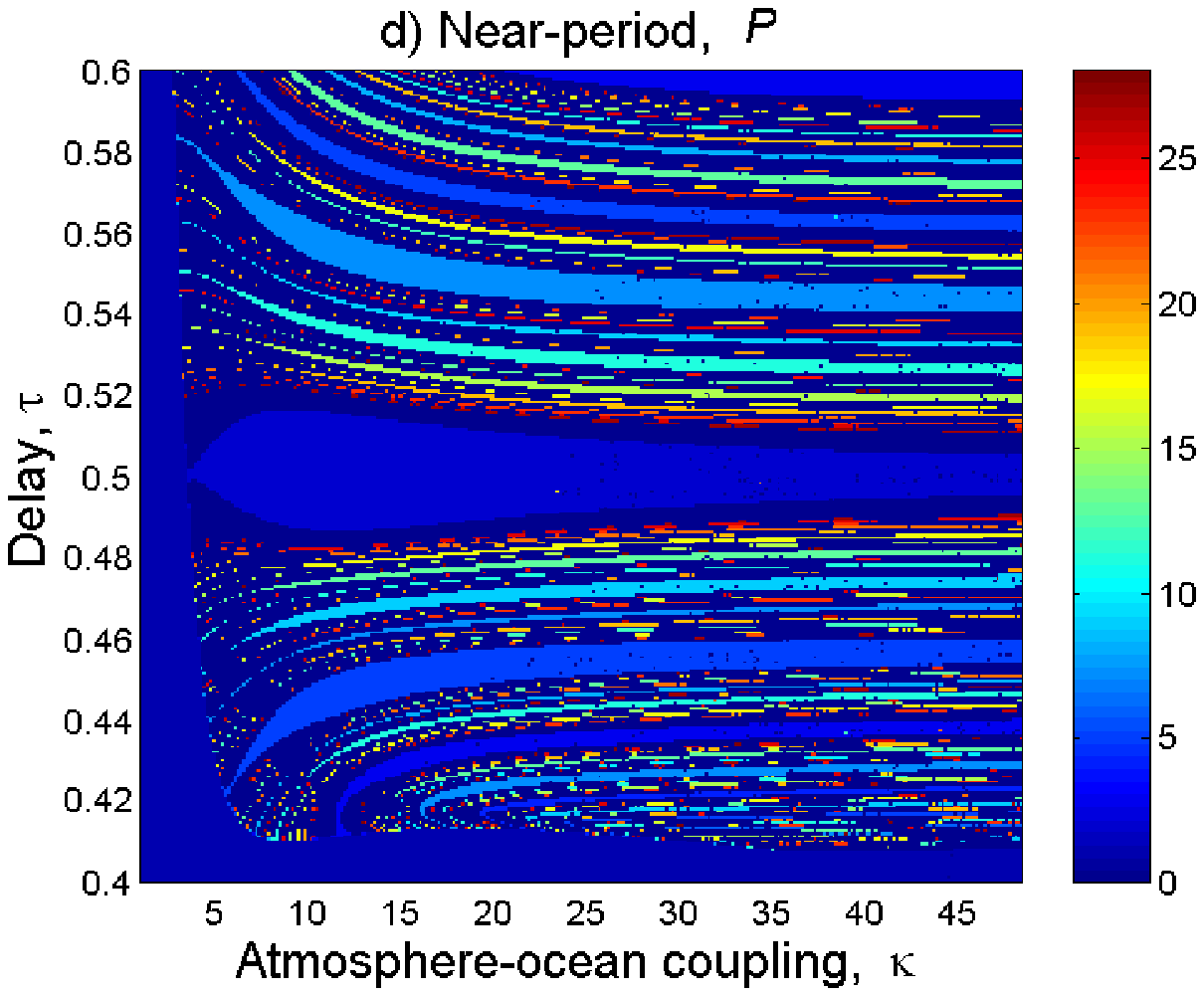}
\caption{Solution statistics plots for $b=1$, as a function
of coupling parameter $\kappa$ and delay $\tau$.
Same panels as in Fig.~\ref{stat_map_a}.}
\label{stat_map_b}
\end{figure*}

Figure~\ref{stat_map_b} shows the same plots of solution statistics 
over a rectangle of the plane $(\kappa,\,\tau)$, for a fixed value $b=1$;
it illustrates the onset and development of instabilities as
a function of the coupling parameter $\kappa$.
Comments similar to those above, which concern the behavior of individual 
maps and the connections between them, apply to this set of maps as well. 

\begin{figure*}[t]
\vspace*{2mm}
\centering\includegraphics[width=16cm]{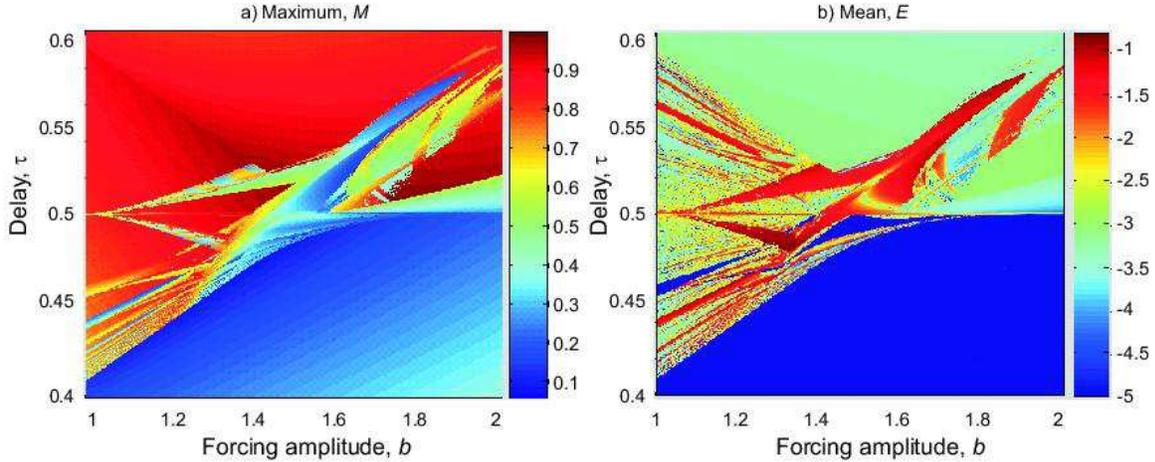}
\caption{Solution statistics plots for $\kappa=100$:
a) maximum map, $M(b,\,\tau)$; and
b) mean map, $\log |E(b,\,\tau)|$.}
\label{stat_map_c}
\end{figure*}

Figure~\ref{stat_map_c} shows the maximum and mean maps for $\kappa=100$. 
The increase in the atmosphere-ocean coupling, with its associated nonlinearity, 
amplifies the model's instabilities and leads to more complex dynamics that
is quite chaotic.
The rigorous verification of the chaotic properties is left, however,
for future work. 

Figure~\ref{fig_instab} shows three examples of change in solution behavior 
across the neutral curve that separates smooth from rougher behavior.
In all three cases shown, one trajectory (dashed line) has period one and lies within
the smooth part of the parameter space, in the immediate vicinity of the
neutral curve, while the other trajectory (solid line) corresponds to a
point in parameter space that is quite close to the first one but on
the other side of the neutral curve. 

Panel (a) illustrates transition to quasiperiodic behavior with a ``carrier wave''
of period near 8, alternating smoothly 3 warm and 3 cold events, separated by
one ``normal'' year.
In panel (b) we see single large El Ni\~nos alternating with single large
La Ni\~nas, separated by 3 normal years.
Panel (c) exhibits interdecadal variability, like in Fig.~2e, except
that the transitions between warm and cold ``decades'' is sharper here,
and the spells of El Ni\~nos and La Ni\~nas even longer (a dozen years
here {\it vs.} roughly ten there). 

\begin{figure}[t]
\vspace*{2mm}
\centering\includegraphics[width=8.3cm]{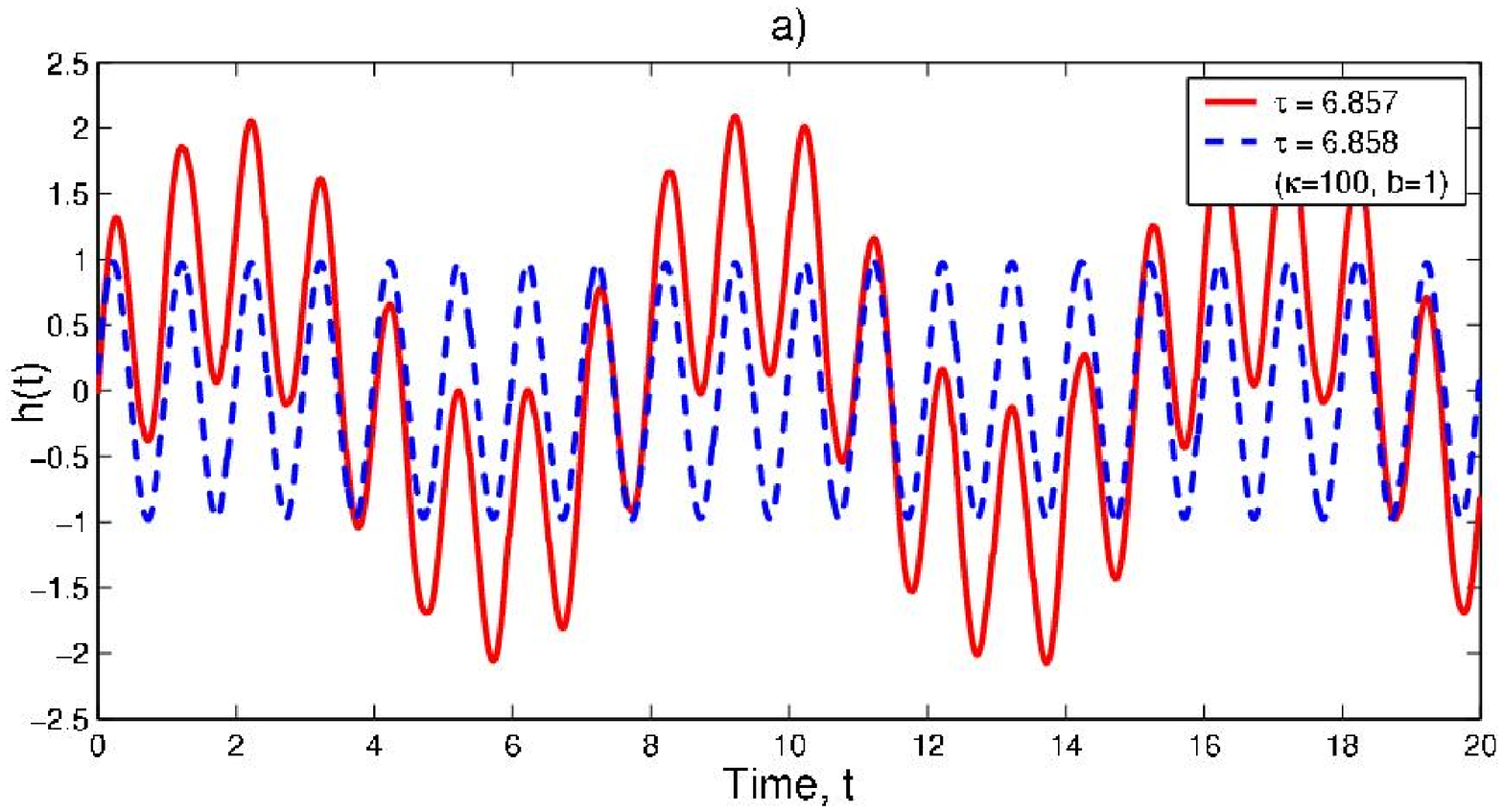}
\centering\includegraphics[width=8.3cm]{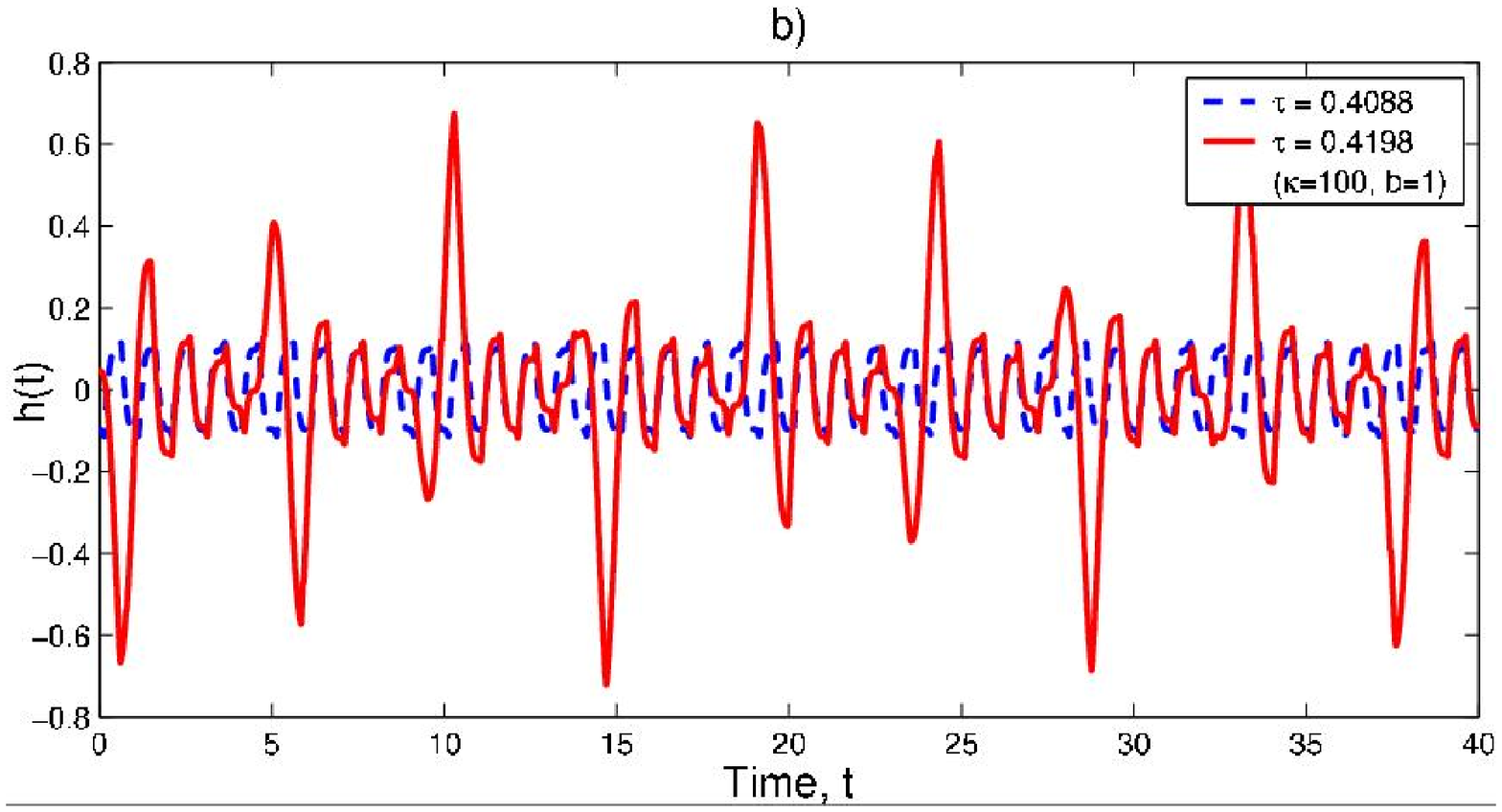}
\centering\includegraphics[width=8.3cm]{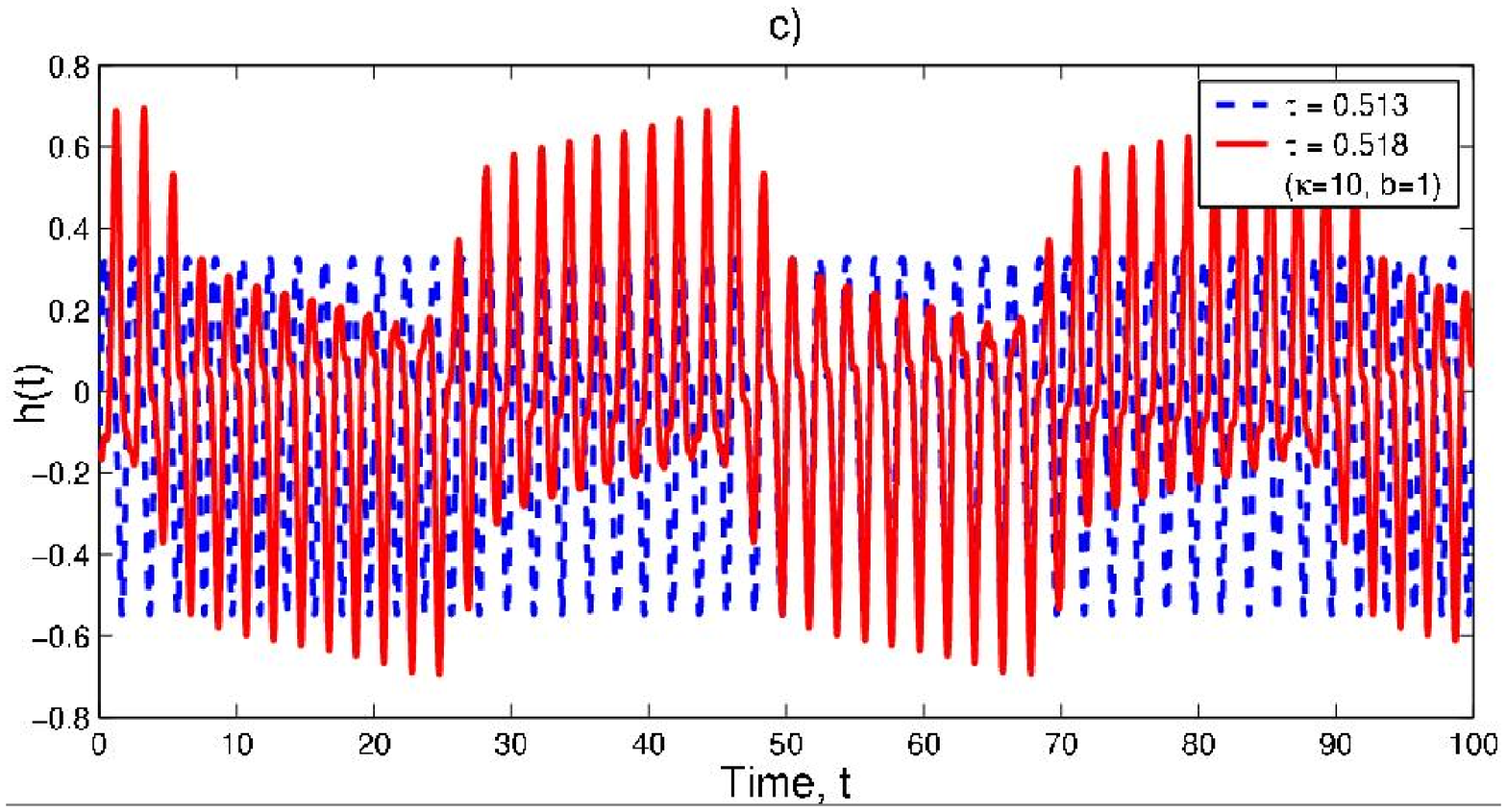}
\caption{Examples of instabilities across the neutral curve
that separates smooth from rough behavior; at $b=1$;
see Fig.~\ref{mmap}.
The dashed blue curve shows a period-1 trajectory on the smooth side of
the neutral curve; the solid red curve shows a trajectory immediately 
across the neutral curve, in the rough-solution domain.
} 
\label{fig_instab}
\end{figure}

\conclusions[Discussion]
\label{discussion}
We have considered a toy model for ENSO variability that is governed by 
a delay differential equation (DDE) with a single, fixed delay and periodic 
forcing.
Thus, we follow a line of research pioneered by Suarez and Schopf
(\citeyear{SS88}), Battisti and Hirst (\citeyear{BH89}), and Tziperman {\it et al.}
(\citeyear{Tzi+94}), who have shown that DDE models can 
effectively capture complex phenomena found in much more detailed ENSO models,
as well as in observational data sets.
DDE models are very simple and, at the same time, exhibit
rich and complex behavior.
Stability and bifurcation analysis for such models can be carried out 
analytically only to some extent, but numerical 
methods are being actively developed \citep{Baker,BPW,ELS},
and we have not yet taken full advantage here of either approach.
 
To initiate stability and bifurcation analysis of ENSO-related 
DDE models, we started here with a descriptive numerical exploration
of Eqs.~\eqref{ivp1}-\eqref{ivp2} over a wide range of physically relevant
parameter values.
We studied parameter dependence of
various trajectory statistics, and report the existence of a large domain in
parameter space where the statistics maps are strikingly discontinuous.
The local continuous-dependence theorem (Proposition \ref{prop}) suggests, at least, 
that the reported discontinuities in global solutions point to the existence of 
unstable solutions of Eqs.~\eqref{ivp1}-\eqref{ivp2};
the complex discontinuity patterns (see Figs.~\ref{stat_map_a} and \ref{stat_map_b})
lead us to suspect the presence of a rich family of unstable solutions that underlie
a complicated attractor.
It would be interesting to study in greater detail the unstable solutions
and understand their role in shaping the system dynamics.

Summarizing the model results in terms of their relevance for ENSO dynamics,
we emphasize the following observations.
A simple DDE model \eqref{ivp1}-\eqref{ivp2} with a single delay reproduces 
the Devil's staircase 
scenario documented in other ENSO models, including ICMs and GCMs, 
as well as in observations \citep{JNG94,JNG96,Tzi+94,Tzi+95,GR00}.
The model illustrates, in simplest possible terms, 
the role of the distinct parameters: strength of seasonal 
forcing $b$ vs. ocean-atmosphere coupling $\kappa$ and transit time $\tau$
of oceanic waves across the Tropical Pacific.  
We find spontaneous transitions in mean thermocline depth, and hence in sea 
surface temperature (SST), as well as in extreme annual values that occur for purely 
periodic, seasonal forcing.
The model generates intraseasonal oscillations of various periods 
and amplitudes, as well as interdecadal variability.
The former result might suggest that Madden-Julian oscillations \citep{MJ71,MJ72,MJ94}
and westerly wind bursts \citep{Geb+07,HG88,Ver98,Del+93} in the Tropical Pacific are affected 
by ENSO's interannual modes at least as much as they affect them in turn.
The latter result likewise suggests that interdecadal variability in the
extratropical, thermohaline circulation \citep{Dij06,DG05} might also interfere
constructively with ENSO's intrinsic variability on this time scale. 
 
A sharp neutral curve in the ($b-\tau$) plane separates smooth parameter 
dependence in the solutions' map of ``metrics'' \citep{Tay01,Fug+03} from ``rough'' behavior.
We expect such separation between regions of smooth and rough dependence 
of solution metrics on
parameters in much more detailed and realistic models, where it is 
harder to describe its causes as completely.

Finally, it appears that, even in as simple a model as our DDE, the mean,
extrema and periodicity of solutions can change (a) spontaneously, without
any change in the external forcing; and (b) one of these characteristics
can change considerably, while others change but very little.
Furthermore, certain parts of parameter space involve only small and smooth
changes, while others involve large and sudden ones.
It is quite conceivable that such behavior might arise in intermediate 
climate models \citep{JNG96,Neel+94,Neel+98} and GCMs \citep{Mur+04,Sta+05}.

\appendix
\section{\\ \\ \hspace*{-7mm} A simple example of DDE complexity}
\label{burst}
In his classical book \citep{Hale} on functional differential equations, 
Jack Hale remarks that systematic study of differential equations with 
dependence on the past started with the work of Volterra on predator-prey
models and viscoelasticity at the beginning of the 20th century.
DDEs have thus been actively studied and applied for almost a century.
Still, they are a relatively new modeling tool when compared to ODEs, and their 
theory and numerical analysis are much less developed than for ODEs. 
To develop the reader's intuition for DDEs, we discuss in this appendix a simple 
autonomous ODE and the corresponding 
DDE obtained by introducing a fixed time delay; our goal is to illustrate
how this apparently innocuous modification complicates the solution set
of the equation and renders its analytical and numerical study more elaborate.  
  
We start with the linear, scalar ODE
\be
\dot x(t) = \alpha\,x(t).
\label{ode}
\ee
Assuming a solution of the form $x=c\,e^{\lambda\,t}$, we substitute it in
\eqref{ode} to find its {\it characteristic equation}
\[\lambda = \alpha.\]
This equation allows us to find all possible functions of the given form 
that satisfy \eqref{ode}. 
Clearly, fixing an initial condition $x(0)=x_0$ leaves only one solution 
with $c=x_0$.
This example illustrates an important general property of autonomous ODEs:
their characteristic equations are polynomials in $\lambda$ and thus have a 
finite set of (complex-valued) solutions that can be easily found.
As a result, the finite set of solutions to \eqref{ode} can also be
described easily.

Let us introduce now a delay $\tau$ into Eq.~\eqref{ode}:
\be
\dot x(t) = \alpha\,x(t-\tau).
\label{dde1}
\ee
This modification implies that it takes some finite time $\tau$ for 
changes in the model state $x(t)$ to affect its rate of 
change $\dot x(t)$.
Such an assumption makes sense in many applications, with numerous 
specific examples given in the Introduction to Hale (\citeyear{Hale}), 
and in Kolmanovskii and Nosov (\citeyear{KN}).
Proceeding as before, we assume a solution
of the form $x=c\,e^{\lambda\,t}$ and obtain the characteristic
equation
\be
\lambda\,e^{\lambda\,\tau}=\alpha.
\label{ch}
\ee
The fact that not all exponential terms cancel out in \eqref{ch}
changes dramatically the solution set of this characteristic equation.
It can be shown that \eqref{ch} has an infinite number 
of complex solutions; hence there exists an infinite number of functions
that satisfy Eq.~\eqref{dde1} \citep{Hale,Falbo}.
A general solution to \eqref{dde1} is given by \citep{Falbo}:
\begin{eqnarray}
x(t)&=&C_0\,e^{-t/\tau}
+C_1\,e^{r_0\,t}+C_2\,e^{r_1\,t}+C_3\,e^{r\,t}\nonumber\\
&+&\sum_{k=1}^{\infty}e^{p_k\,t}
\left[C_{1k}\,\cos(q_k\,t)+C_{2k}\,\sin(q_k\,t)\right].
\end{eqnarray}
Here $p_k\pm i\,q_k$ are complex solutions of \eqref{ch},
$C_{1k}$ and $C_{2k}$ are arbitrary constants, and
$C_0, C_1, C_2, C_3$ depend on the values of $\alpha$
and $\tau$:
for $\alpha<-1/(\tau\,e)$, $C_i=0$, $i=0,1,2,3$;
for $\alpha=-1/(\tau\,e)$, $C_0$ is arbitrary, and
$C_i=0$, $i=1,2,3$;
for $-1/(\tau\,e)<\alpha<0$, $C_0=C_3=0$, $C_1, C_2$
are arbitrary, and $r_0$, $r_1$ are the real roots of
\eqref{ch};
finally for $\alpha>0$, $C_i=0$ for $i=0,1,2$,
$C_3$ is arbitrary, and $r$ is the only real root
of \eqref{ch}.
Accordingly, this simple autonomous linear DDE may have
increasing, decreasing, and oscillating solutions.

The ``burst'' of complexity in the solution set of \eqref{dde1}
compared to that of \eqref{ode} is pervasive in the DDE realm.
Solving characteristic equations arising from DDEs is typically 
quite involved and requires usually some general results on equations 
that mix polynomial and exponential terms (see, {\it e.g.,} 
Appendix A in \citet{Hale}).
Stability and bifurcation studies for DDEs face a similar
``burst'' of complexity, compared to ODEs.

Numerical exploration of DDEs is also considerably more complicated
than for ODEs.
The question of discontinuities in the solutions' derivatives is
crucial:
Because general numerical methods for both ODEs and DDEs are intended
for problems with solutions that have several continuous
derivatives, discontinuities in low-order derivatives (up to
the order of the integration method) require special attention.
Such discontinuities are not all that rare for ODEs, but they are
almost always present for DDEs, since the first derivative of
the history function at the initial point is almost always different 
from the first derivative of the solution there.
Moreover, discontinuities are a much more serious matter for DDEs than
they are for ODEs because they propagate:
the smoothness of the derivative $\dot x$ at the current time
$t$ depends on the smoothness of the solution $x$ at past
times in the interval $t-\tau$.

This difficulty may be illustrated using our very simple example
\begin{equation}
 \dot x(t) = x(t-1),
\label{ex}
\end{equation}
where we set, without loss of generality, $\tau=\alpha=1$.
For this equation $x^{(k+1)}(t) = x^{(k)}(t-1)$, where
$x^{(k)}$ denotes the $k$-th derivative.
In general, if there is a discontinuity of order $k$ at a time $t^*$, 
meaning that $x^{(k)}$ has a jump at $t = t^*$, then as $t$ crosses 
the value $t^* + 1$, there is a discontinuity in
$x^{(k+1)}$ because of the term $x(t - 1)$.
With multiple delays $\tau_i$, a discontinuity at time $t^*$
propagates to the times
\[t^* + \tau_1,\, t^* + \tau_2,\, \ldots,\, t^* + \tau_k,\]
and each of these discontinuities in turn propagates.
If there is a discontinuity of order $k$ at time $t^*$, the
discontinuity at each of the times $t^* + \tau_j$ is at least 
of order $k+1$.
Because the effect of a delay appears in a higher-order derivative, 
the solution does become smoother as the integration proceeds.
This ``generalized smoothing'' proves to be quite important to
the numerical solution of DDEs.

To illustrate these statements,
suppose we wish to solve Eq.~\eqref{ex} with history $x(t)\equiv 1$ for
$t \le 0$.
On the interval $0 \le t \le 1$ the solution is $x(t-1) = 1$ because $t-1 \le 0$.
Thus, the DDE on this interval reduces to the ODE $\dot x(t)=1$
with initial value $x(0) = 1$.
We solve this problem to obtain $x(t)=t+1$ for $0 \le t \le 1$.
Notice that this solution exhibits a typical discontinuity
in its first derivative at $t=0$ because it is zero to the left
of the origin and unity to the right.
Now that we know the solution for $t \le 1$, we can reduce the DDE
on the interval $1 \le t \le 2$ to an ODE $\dot x= (t-1)+1 = t$
with initial value $x(1)=2$ and solve this problem to find that
$x(t)= 0.5 t^2 + 1.5$ on this interval.
The first derivative is continuous at $t=1$, but there is a
discontinuity in the second derivative.
In general, the solution of the DDE on the interval $[k,k+1]$ is
a polynomial of degree $k+1$ and the solution has a
discontinuity of order $k+1$ at time $t=k$.

In order for a solver to account for these specific features of DDEs and to
solve them efficiently, accurately and reliably, there must be a great
deal of care taken ``under the hood'' of the solver.
Discontinuities need be tracked only up to the order of the
integration method, since higher-order discontinuities do not affect
the performance of the solver, {\it i.e.} its  procedures for
error estimation and step size control.
Our solver of choice, \texttt{dde\_solver}, tracks discontinuities
explicitly and includes them as integration grid points, in order
to avoid interpolating across them.
For problems with constant delays, it is possible to build the
necessary ``discontinuity tree'' in advance and to step exactly
to each point in the tree.

For problems with state-dependent delays, the discontinuity times
are not known in advance.
In this case, \texttt{dde\_solver} tracks discontinuities using
root finding, in conjunction with the primary integration method's
underlying polynomial interpolants, to locate the discontinuities
and restart the integration at each such point.
Some other available solvers handle discontinuity propagation 
differently; but the best solvers do take special
precautions of one kind or another, since ignoring discontinuities
can significantly affect the reliability of a DDE solver.
We refer in this context also to several distinct approaches 
to the numerical solution of Boolean delay equations
\citep{DG84,GM85,SG01,ZKG03}.

\section{\\ \\ \hspace*{-7mm} Proof of Proposition~\ref{prop}}
\label{proof}
Consider the IVP \eqref{ivp1}-\eqref{ivp2}, with the rhs of the DDE
\eqref{ivp1} denoted by
\[\cF(t,h(\cdot))=-\tanh\left[\kappa\,h(\cdot)\right]+b\,\cos(2\,\pi\,t).\]

{\it Existence} of the solution to this problem on $[0,T]$ for some $T>0$ readily 
follows from the continuity of $\cF(t,h)$ and the general existence theorem 
for DDEs (\citet{Hale}, Theorem~2.1, p. 41).
Moreover, Nussbaum (\citet{Nuss}, Theorem~1, p. 3) remarks that if there
exist constants $A$ and $B$ such that 
$\parallel \cF(t,h)\parallel \le A\parallel h\parallel+B$
for all $(t,h)\in\mathbb{R}\times X$ then one can choose $T=\infty$. 
Since $|\tanh(\kappa\,z)|\le\kappa\,|z|$, the choice of $A=\kappa$ and 
$B=b$ ensures the solution's existence on $[0,\infty)$.

{\it Uniqueness} could be derived from the Lipschitz property of $\cF(t,h)$
in $h$ and the general uniqueness theorem (\citet{Hale}, Theorem~2.3, p. 42).
However, for our system the uniqueness can be established in a simpler way.
Indeed, assume that $x(t)$ and $y(t)$ are solutions of \eqref{ivp1}-\eqref{ivp2}
on $[0,\,T]$, with rhs $\cF(t,h(t-\tau))$ and the initial condition $\phi(t)$.
Then, for $0< t \le \tau$,
\begin{eqnarray}
\lefteqn{x(t)-y(t)=\int_0^t \left[\vphantom{I^I}\cF(s,x(s-\tau)-\cF(s,y(s-\tau)\right]\,ds}\nonumber\\
&=&\int_0^t \left(\vphantom{I^I}\tanh[\kappa\,y(s-\tau)]-\tanh[\kappa\,x(s-\tau)]\right)\,ds\nonumber\\
&=&\int_{-\tau}^{t-\tau}
\left(\vphantom{I^I}\tanh[\kappa\,y(u)]-\tanh[\kappa\,x(u)]\right)\,du\equiv 0.
\end{eqnarray}
Thus, the solutions $x$ and $y$ are identical up to $t=\tau>0$.
The uniqueness is proven by successively advancing in time by intervals
of length $\tau$.

{\it Continuous dependence} on initial conditions and the rhs of \eqref{ivp1}
for any finite $T$ 
follows from the existence and uniqueness and the general continuous dependence 
theorem (\citet{Hale}, Theorem~2.2, p. 41).
To show the continuous dependence on the delay $\tau$, also for finite $T$, we consider
the sequence $\tau_k\to\tau$, $k\to\infty$, and for any $\tau_k$
introduce the time scale change: $s_k=t\,\tau/\tau_k$.
Then, 
\[h(t-\tau_k)=h\left(\frac{\tau_k}{\tau}(s_k-\tau)\right)\equiv h_k(s_k-\tau),\]
and one finds
\[\frac{d}{ds_k} h_k(s_k) =-\frac{\tau_k}{\tau}\,A_{\kappa}\left[h_k(s_k-\tau)\right]
+b\,\cos\left(2\,\pi\,\frac{s_k\,\tau_k}{\tau}\right).\]
Clearly, the rhs of this system converges to $\cF$ as $k\to\infty$.
This shows that a small change in the delay $\tau$ can be considered as
a small change of the rhs of the Eq.~\eqref{ivp1} with the same delay.
Hence, continuous dependence on the rhs implies continuous dependence
on the delay.

\section{\\ \\ \hspace*{-7mm} DDE solvers}
\label{solver}

\subsection{Our solver of choice: \texttt{dde\_solver}}
The DDE solver \texttt{dde\_solver}
\citep{F90} was used to perform the numerical experiments reported
in this paper;
\texttt{dde\_solver} is a Fortran 90/95 extension of its
Fortran 77 predecessor \texttt{dklag6} \citep{dklag6}.
Both \texttt{dde\_solver} and \texttt{dklag6} implement a
(5,6) pair of continuously embedded, explicit Runge-Kutta-Sarafyan
methods.
We refer to \citep{dklag6} for the coefficients and precise details
of the methods used, and to \citep{lfs} for a discussion of
continuously embedded Runge-Kutta methods.
Both methods in the pair are based on piecewise-polynomial
approximants, which are used for error estimation and step size
selection, to handle the necessary interpolations for delayed
solution values, and to track derivative discontinuities that
are propagated by the system delays, while
the sixth-order method is used to perform the actual integration.

As discussed in \citep{F90}, \texttt{dde\_solver} was designed to
solve systems of DDEs with state-dependent delays in as ``user-friendly''
a fashion as possible, while at the same time retaining and extending
the solution capabilities of \texttt{dklag6}.
Our solver was also designed so that usage approaches the convenience of
the \textsc{Matlab} (\citeyear{matlab}) DDE solvers  \texttt{dde23} and
\texttt{ddesd}.
For example, storage management is handled automatically by the solver,
thus relieving the user of the necessity to supply work arrays whose
sizes are not known in advance.
Several options are available for supplying necessary information
about the problem and for dealing with its special characteristics.
All options have carefully chosen defaults that can be changed by the
user.
These include the ability to supply vectors or functions to define
the delays and the initial solution history, the ability to specify
points corresponding to known derivative or solution discontinuities,
tracking delay-induced derivative discontinuities, the ability to
cope with small delays, the ability to handle state-dependent events
({\it e.g.}, times at which it is desirable to make qualitative changes or
parameter changes in the underlying system of DDEs), and the ability
to solve so-called neutral DDEs, which contain delayed derivatives.
The solver builds and returns a Fortran 90 solution structure that
may be used for various tasks, {\it e.g.}, for plotting purposes.
An interpolation module uses this structure, for example, to perform
additional interpolations requested by the user.

The numerical studies in this paper led us to incorporate several
new options in \texttt{dde\_solver}.
By default, the solver retains the entire solution history.
The numerical simulations in this paper, though, require very long
integration intervals, leading to prohibitive storage requirements.
We added, therefore, an option to have the solver trim points from 
the solution history queue that precede the largest delay.
A related option was added to allow the user to provide a
module to process solution information before it is discarded, thus
retaining the ability for user interpolation.

The \texttt{dde\_solver} with these added options is available at
\url{http://www.radford.edu/~thompson/ffddes/}.
In addition to the solver, a variety of example programs can also be
found there; they may be used as convenient templates for other
problems.
This solver is a Fortran 90/95 compliant self-contained module with no
restrictions on its use.
In particular, it is not compiler dependent and has been used successfully with most 
of the available F90/F95 compilers, including, for example, g95, 
Lahey LF90, Lahey-Fujitsu LF95, Salford FTN95, SUNf95, and Compaq.

\subsection{Other DDE solvers}
Other capable DDE software is available.
Some noteworthy solvers include
\texttt{archi}, \texttt{ddverk}, \texttt{dde23}, and \texttt{ddesd}.
Like \texttt{dde\_solver}, the first three of these solvers implement pairs
of continuously embedded, explicit Runge-Kutta methods.
Thus, \texttt{archi} \citep{archi} is based on a (4,5) pair, and
allows the user to specify either extrapolation or iterative
evaluation of implicit formulas.
The solver \texttt{ddverk} \citep{ddverk} is based on a (5,6) pair
and it handles small and vanishing delays iteratively.
Defect error control is used to detect suspected derivative
discontinuities, locate them and use special interpolants
when stepping over them.
The two solvers \texttt{dde23} \citep{dde23} and \texttt{ddesd} \citep{ddesd} are
available in the \textsc{Matlab} problem solving environment;
\texttt{dde23} is based on a (2,3) pair and is applicable to DDEs
with constant delays.
It is worth noting that the \texttt{dde23}
user interface led to many of the design decisions used in
\texttt{dde\_solver}.
The solver \texttt{ddesd} incorporates novel methods based on control
of the solution residual and is intended for systems with
state-dependent delays.
In addition to these solvers, two other popular and well-known tools
include the \texttt{DDE-BIFTOOL} package \citep{ELS} and the
\texttt{XPPAUT} package \citep{xppaut}.
Each of these packages contains a variety of tools that are useful
for analyzing delayed dynamical systems.

\begin{acknowledgements}
We are grateful to our colleagues M. Chekroun, J. C. McWilliams, J. D. Neelin and 
E. Simonnet for many useful discussions and their continuing interest in this work.
The study was supported by DOE Grant DE-FG02-07ER64439 from the Climate Change 
Prediction Program, by NSF Grant ATM-0620838, and by the European Commission's 
No. 12975 (NEST) project ``Extreme Events: Causes and Consequences (E2-C2).''
\end{acknowledgements}


\begin{thebibliography}{99}

\bibitem[Andronov and Pontryagin(1937)]{AP37}
Andronov, A. A. and Pontryagin, L. S.:
Syst\`emes grossiers,  
{Dokl. Akad. Nauk SSSR}, {14}(5), 
247-250, 1937.

\bibitem[Baker(2000)]{Baker}
Baker, C. T. H.:
Retarded differential equations,
{J. Comp. Appl. Math.}, {125}, 309-335, 2000.

\bibitem[Baker {\it et al.}(1995)]{BPW}
Baker, C. T. H., Paul, C. A. H., and Will\'e, D.R.:
A bibliography on the numerical solution of delay differential equations, 
{Numerical Analysis Report}, {269}, 
Manchester Centre for Computational Mathematics, Manchester, England, 1995.

\bibitem[Battisti(1988)]{Batt88}
Battisti, D. S.:
The dynamics and thermodynamics of a warming event in a 
coupled tropical atmosphere/ocean model,
{J. Atmos. Sci.}, {45}, 2889--2919, 1988.

\bibitem[Battisti and Hirst(1989)]{BH89}
Battisti, D. S. and Hirst, A. C.:
Interannual variability in the tropical
atmosphere-ocean system: Influence of the
basic state and ocean geometry,
{J. Atmos. Sci.}, {46}, 1687-1712, 1989.


\bibitem[Bhattacharrya and Ghil(1982)]{BG82}
Bhattacharrya, K., and Ghil, M.:
Internal variability of an energy-balance model
with delayed albedo effects, 
{J. Atmos. Sci.}, {39}, 1747--1773, 1982.

\bibitem[Bjerknes(1969)]{Bjer69}
Bjerknes, J.:
Atmospheric teleconnections from the equatorial Pacific,
Mon. Wea. Rev., 97, 163--172, 1969.

\bibitem[Bodnar(2004)]{Bod04}
Bodnar, M.:
On the differences and similarities of the first-order
delay and ordinary differential equations,
{J. Math. Anal. Appl.}, {300}, 172-188, 2004.

\bibitem[Cao(1996)]{Cao}
Cao, Y.: 
Uniqueness of periodic solution
for differential delay equations,
{J. Diff. Eq.}, {128}, 46-57, 1996.


\bibitem[Chang {\it et al.}(1994)]{Cha94}
Chang, P., Wang, B., Li, T. and Ji, L.:
Interactions between the seasonal cycle and the Southern Oscillation:
Frequency entrainment and chaos in intermediate coupled ocean-atmosphere
model,
Geophys. Res. Lett., {21}, 2817--2820, 1994.

\bibitem[Chang {\it et al.}(1995)]{Cha95}
Chang, P., Ji, L., Wang, B., and Li, T.:
Interactions between the seasonal cycle and El Ni\~no -
Southern Oscillation in an intermediate coupled 
ocean-atmosphere model,
{J. Atmos. Sci.}, {52}, 2353--2372, 1995.

\bibitem[Chow and Walter(1988)]{CW88}
Chow, S.-N., and Walter, H. O.:
Characteristic multipliers and stability of periodic solutions
of $\dot x=g\left(x(t-1)\right)$,
{Trans. Amer. Math. Soc.}, {307}, 127-142, 1988.

\bibitem[Corwin {\it et al.}(1997)]{dklag6} 
Corwin, S.P., Sarafyan, D., and Thompson, S.:
DKLAG6: a code based on continuously imbedded sixth order
Runge-Kutta methods for the solution of state dependent
functional differential equations,
{Appl. Numer. Math.}, {24}, 319-333, 1997.

\bibitem[Dee and Ghil(1984)]{DG84}
Dee D. and Ghil, M.:
Boolean difference equations, I: Formulation and dynamic behavior,
{SIAM J. Appl. Math.}, {44}, 111-126, 1984.

\bibitem[Delcroix {\it et al.}(1993)]{Del+93}
Delcroix, T., Eldin, G., McPhaden, M., and Morli\`ere, A.:
Effects of westerly wind bursts upon the western 
equatorial Pacific Ocean, February-April 1991,
{J. Geophys. Res.}, {98} (C9), 16~379-16~385, 1993.

\bibitem[Diaz and Markgraf(1992)]{Diaz92}
Diaz, H. F. and Markgraf, V. (Eds.): 
El Ni\~no: Historical and Paleoclimatic 
Aspects of the Southern Oscillation,
Cambridge Univ. Press, New York, 1993.

\bibitem[Dijkstra(2005)]{Dij06}
Dijkstra, H. A.:
{Nonlinear Physical Oceanography: A Dynamical Systems Approach 
to the Large Scale Ocean Circulation and El Ni\~no}, 
2nd edition, Springer-Verlag, 2005.

\bibitem[Dijkstra and Ghil(2005)]{DG05}
Dijkstra, H.A. and Ghil, M.:
Low-frequency variability of the ocean circulation: 
a dynamical systems approach, 
{Rev. Geophys.}, {43}, RG3002, doi:10.1029/2002RG000122, 2005. 

\bibitem[Engelborghs {\it et al.}(2001)]{ELS}
Engelborghs, K., Luzyanina, T., and Samaey, G.:
DDE-BIFTOOL v. 2.00: a Matlab package for numerical bifurcation analysis
of delay differential equations,
Report TW, vd330, Department of Computer Science, 
K.U. Leuven, Leuven, Belgium, 2005. 
Available at
\url{http://www.cs.kuleuven.ac.be/cwis/research/twr/research/software/delay/ddebiftool.shtml}.

\bibitem[Enright and Hayashi(1997)]{ddverk} 
Enright, W.H. and Hayashi, H.:
A delay differential equation solver based on a continuous Runge-Kutta
method with defect control, 
{Numer. Alg.}, {16}, 349-364, 1997.

\bibitem[Ermentrout(2002)]{xppaut}
Ermentrout, B.:
{Simulating, Analyzing, and Animating Dynamical Systems:
A Guide to XPPAUT for Researchers and Students}, SIAM,
Philadelphia, USA, 2002.

\bibitem[Falbo(1995)]{Falbo}
Falbo, C. E.:
Analytical and numerical solutions to the
delay differential equation $\dot y=\alpha\,y(t-\delta)$,
{Proc. Joint Mtg. California sections of the MAA}, 
San Luis Obispo, CA, October 21, 1995.
Available at:
\url{http://www.sonoma.edu/math/faculty/falbo/pag1dde.html}.

\bibitem[Fuglestvedt {\it et al.}(2003)]{Fug+03}
Fuglestvedt, J.S., Berntsen, T. K., Godal, O., Sausen, R., 
Shine, K. P., and Skodvin, T.: 
Metrics of climate change: Assessing radiative forcing and emission indices, 
{Climatic Change}, {58}, 267-331, 2003.

\bibitem[Gebbie {\it et al.}(2007)]{Geb+07}
Gebbie, G., Eisenman, I., Wittenberg, A., and Tziperman, E.:
Modulation of westerly wind bursts by sea surface temperature:
A semistochastic feedback for ENSO,
{J. Atmos. Sci.}, {64}, 3281-3295, 2007. 


\bibitem[Ghil {\it et al.}(2002)]{Ghil+02}
Ghil, M., Allen, M. R., Dettinger, M. D., Ide, K., Kondrashov, D., 
Mann, M. E., Robertson, A. W., Saunders, A., Tian, Y., 
Varadi, F., and Yiou, P: 
Advanced spectral methods for climatic time series,
{Rev. Geophys.}, {40} (1), Art. No. 1003, 2002.

\bibitem[Ghil and Childress(1987)]{GC87}
Ghil, M. and Childress, S.:
{Topics in Geophysical Fluid Dynamics:
Atmospheric Dynamics, Dynamo Theory, and 
Climate Dynamics,}
Springer, Verlag, 1987.

\bibitem[Ghil and Mullhaupt(1985)]{GM85}
Ghil, M. and Mullhaupt, A. P.:
Boolean delay equations. II: Periodic and aperiodic solutions,
{J. Stat. Phys.}, {41}, 125-173, 1985.


\bibitem[Ghil and Robertson(2000)]{GR00}
Ghil, M. and  Robertson, A. W.:
Solving problems with GCMs: 
General circulation models and 
their role in the climate modeling hierarchy,
In D.  Randall  (Ed.) 
{General Circulation Model Development: 
Past, Present and Future},
Academic Press, San Diego, 285--325, 2000.

\bibitem[Glantz {\it et al.}(1991)]{Glantz+91}
Glantz, M. H., Katz, R. W., and Nicholls, N. (Eds.):   
Teleconnections Linking Worldwide Climate Anomalies, 
Cambridge Univ. Press, New York, 545 pp, 1991.

\bibitem[Hale(1977)]{Hale}
Hale, J.:
{Theory of Functional Differential Equations,}
Springer-Verlag, New-York, 1977.

\bibitem[Harrison and Giese(1988)]{HG88}
Harrison, D. E. and Giese, B.:
Remote westerly wind forcing of the eastern equatorial Pacific; 
some model results,
{Geophys. Res. Lett.}, {15}, 804-807, 1988.

\bibitem[Jiang {\it et al.}(1995)]{Jiang+95}
Jiang, N., Neelin, J. D., and Ghil, M.:
Quasi-quadrennial and quasi-biennial variability in the equatorial Pacific,
{Clim. Dyn.}, {12}, 101-112, 1995.

\bibitem[Jin {\it et al.}(1994)]{JNG94}
Jin, F.-f., Neelin, J. D., and Ghil, M.:
El Ni\~no on the Devil's Staircase: 
Annual subharmonic steps to chaos,
{Science}, {264}, 70--72, 1994.

\bibitem[Jin {\it et al.}(1996)]{JNG96}
Jin, F.-f., Neelin, J. D., and Ghil, M.:
El Ni\~no/Southern Oscillation and the annual cycle:
Subharmonic frequency locking and aperiodicity,
{Physica D}, {98}, 442--465, 1996.

\bibitem[Katok and Hasselblatt(1995)]{KH}
Katok, A. and Hasselblatt, B.:
{Introduction to the Modern Theory of Dynamical Systems}, 
Cambridge University Press, 802 pp, 1995.

\bibitem[Kolmanovskii and Nosov(1986)]{KN}
Kolmanovskii, V. B. and Nosov, V. R.:
{Stability of Functional Differential Equations,}
Academic Press, 1986.

\bibitem[Latif {\it et al.}(1994)]{Latif94}
Latif, M., Barnett, T. P., Fl\"ugel, M., Graham, N. E., 
Xu, J.-S., and Zebiak, S. E.: 
A review of ENSO prediction studies,
Clim. Dyn., 9, 167--179, 1994.

\bibitem[Madden and Julian(1971)]{MJ71}
Madden, R. A. and Julian, P. R.:
Description of a 40--50 day oscillation in the zonal wind
in the tropical Pacific,
{J. Atmos. Sci.}, {28}, 702-708, 1971.

\bibitem[Madden and Julian(1972)]{MJ72}
Madden, R. A. and Julian, P. R.:
Description of global-scale circulation cells in the tropics
with a 40--50 day period,
{J. Atmos. Sci.}, {29}, 1109-1123, 1972.

\bibitem[Madden and Julian(1994)]{MJ94}
Madden, R. A. and Julian, P. R.:
Observations of the 40--50-day tropical oscillation --- A review,
{Mon. Wea. Rev.}, {122}(5), 814-37, 1994.

\bibitem[Matlab(2007)]{matlab} \textsc{Matlab} {7.4}: 
The MathWorks, Inc., 3 Apple Hill Dr., Natick, MA 01760, 2007.

\bibitem[Munnich {\it et al.}(1991)]{MCZ91}
Munnich, M., Cane, M., and Zebiak, S. E.:
A study of self-excited oscillations of the
tropical ocean-atmosphere system. Part II: Nonlinear cases,
{J. Atmos. Sci.}, {48} (10), 1238-1248, 1991.

\bibitem[Murphy {\it et al.}(2004)]{Mur+04}
Murphy J. M., Sexton, D. M. H., Barnett, D. N., Jones, G. S., 
Webb, M. J., Collins, M.:
Quantification of modelling uncertainties in a large ensemble 
of climate change simulations,
{Nature}, {430}(7001), 768-772, 2004.

\bibitem[Neelin {\it et al.}(1994)]{Neel+94}
Neelin, J. D., Latif, M., and Jin, F.-F.:
Dynamics of coupled ocean-atmosphere models: the tropical problem, 
{Ann. Rev. Fluid Mech.}, {26}, 617-659, 1994.

\bibitem[Neelin {\it et al.}(1998)]{Neel+98}
Neelin, J. D., Battisti, D. S., Hirst, A. C., Jin, F.-F., 
Wakata, Y., Yamagata, T., and Zebiak, S.:
ENSO Theory,
{J. Geophys. Res.}, {103}(C7), 14261-14290, 1998.

\bibitem[Nussbaum(1979)]{Nus79}
Nussbaum, R. D.:
Uniqueness and nonuniqueness for periodic solutions
of $\dot x(t)=-g\left(x(t-1)\right)$, 
{J. Diff. Eq.}, {34}, 25-54, 1979.


\bibitem[Nussbaum(1998)]{Nuss}
Nussbaum, R. D.: Functional Differential Equations, 1998.
Available at 
\url{http://citeseer.ist.psu.edu/437755.html}.

\bibitem[Paul(1995)]{archi} 
Paul, C. A. H.:
{A user-guide to ARCHI},
Numerical Analysis Report {283},
Mathematics Department, University of Manchester, U.K., 1995.

\bibitem[Philander(1990)]{Phil90}
Philander, S. G. H.:
El Ni\~no, La Ni\~na, and the Southern Oscillation,
Academic Press, San Diego, 1990.

\bibitem[Saunders and Ghil(2001)]{SG01}
Saunders, A. and Ghil, M.:
A Boolean delay equation model of ENSO
variability,
{Physica D}, {160}, 54--78, 2001.

\bibitem[Shampine(1994)]{lfs}
Shampine, L. F.:
{Numerical Solution of Ordinary Differential Equations},
Chapman\&Hall, New York, NY, 1994.

\bibitem[Shampine(2005)]{ddesd}
Shampine, L. F.:
Solving ODEs and DDEs with residual control,
{Appl. Numer. Math.}, {52}, 113-127, 2005.

\bibitem[Shampine and Thompson(2001)]{dde23}
Shampine, L. F. and Thompson, S.:
Solving DDEs in \textsc{Matlab}, 
{Appl. Numer. Math.}, {37}, 441-458, 2001.

\bibitem[Shampine and Thompson(2006)]{F90}
Shampine, L. F. and Thompson, S.:
A friendly Fortran 90 DDE solver,
{Appl. Num. Math.}, {56}, (2-3), 503-516, 2006.

\bibitem[Stainforth {\it et al.}(2005)]{Sta+05}
Stainforth D. A., Aina, T., Christensen, C., {\it et al.}:
Uncertainty in predictions of the climate response to rising 
levels of greenhouse gases, 
{Nature}, {433}(7024), 403-406, 2005.

\bibitem[Suarez and Schopf(1988)]{SS88}
Suarez, M. J. and Schopf, P. S.:
A delayed action oscillator for ENSO,
J. Atmos. Sci, 45, 3283-3287, 1988.

\bibitem[Taylor(2001)]{Tay01}
Taylor, K. E.:
Summarizing multiple aspects of model performance in a single diagram,
{J. Geophys. Res.}, {106}, 7183-7192, 2001.

\bibitem[Tziperman {\it et al.}(1994)]{Tzi+94}
Tziperman, E., Stone, L., Cane, M., and Jarosh, H.:  
El Ni\~no chaos:  Overlapping of resonances between 
the seasonal cycle and the Pacific ocean-atmosphere 
oscillator,
{Science}, {264}, 72--74, 1994.

\bibitem[Tziperman {\it et al.}(1995)]{Tzi+95}
Tziperman, E., Cane, M. A., and Zebiak, S. E.:
Irregularity and locking to the seasonal cycle in an 
ENSO prediction model as explained by the quasi-periodicity 
route to chaos,
{J. Atmos. Sci.}, {50}, 293--306, 1995.

\bibitem[Verbickas(1998)]{Ver98}
Verbickas, S.:
Westerly wind bursts in the tropical Pacific,
{Weather}, {53}, 282-284, 1998.

\bibitem[Yanai and Li(1994)]{YL94}
Yanai, M. and Li, C.: 
Interannual variability of the Asian summer monsoon 
and its relationship with ENSO, Eurasian snow cover, 
and heating, 
{Proc. Int. Conf. on Monsoon Variability and Prediction}, 
Trieste, Italy,  9-13 May 1994, 
World Meteorological Organization, 27--34.

\bibitem[Zaliapin {\it et al.}(2003)]{ZKG03}
Zaliapin, I., Keilis-Borok, V., and Ghil, M.:
A Boolean delay equation model of colliding cascades.
Part I: Multiple seismic regimes,
{J. Stat. Phys.}, {111}, 815-837, 2003.

\bibitem[Zebiak and Cane(1987)]{CZ}
Zebiak, S. and Cane, M. A.:
A model El-Ni\~no Southern Oscillation,
{Mon. Wea. Rev.}, {115}, 2262-2278, 1987.


\end{thebibliography}
\end{document}